\documentclass[11pt,a4paper]{article}
\usepackage{jheppub}

\usepackage[utf8]{inputenc}

\usepackage{graphicx}
\graphicspath{ {./figures/} }

\usepackage{amsmath} 
\allowdisplaybreaks
\usepackage{physics}

\usepackage{multirow}

\usepackage{float}
\usepackage{subfig}

\newcommand\RE{\operatorname{Re}} 
 
\newcommand\FS{\alpha_{\text{EM}}} 
\newcommand\FShat{\hat{\alpha}_{\text{EM}}} 
\newcommand\MSbar{$\overline{\text{MS}}$ } 
\newcommand\Veff{V_\text{eff}} 
\newcommand\he[1]{#1^\dagger}
\newcommand\gr[1]{\mathrm{#1}}

\let\vec\mathbf 

\newcommand{\GeV}{{\ensuremath\rm \,GeV}}
\newcommand{\doublet}[2]{ \left( \begin{array}{c}#1 \\ #2 \end{array}\right) }
\newcommand{\be}{\begin{equation}}
\newcommand{\ee}{\end{equation}}
\newcommand{\bea}{\begin{eqnarray}}
\newcommand{\eea}{\end{eqnarray}}

\begin{document}

\title{On the validity of perturbative studies of the electroweak phase transition in the Two Higgs Doublet model}

\preprint{HIP-2019-11/TH \\ \mbox{} \hfill KCL-PH-TH/2019-40}

\author[a]{Kimmo Kainulainen,}
\emailAdd{kimmo.kainulainen@jyu.fi}
\affiliation[a]{Department of Physics, University of Jyväskylä, P.O.~Box 35, FI-40014 University 
of Jyväskylä, Finland}

\author[b]{Venus Keus,}
\emailAdd{venus.keus@helsinki.fi}
\affiliation[b]{Department of Physics and Helsinki Institute of Physics, P.O.~Box 64, FI-00014 University 
of Helsinki, Finland}

\author[b]{Lauri Niemi,}
\emailAdd{lauri.b.niemi@helsinki.fi}

\author[b]{Kari Rummukainen,}
\emailAdd{kari.rummukainen@helsinki.fi}

\author[c]{Tuomas V.I.~Tenkanen}
\emailAdd{tenkanen@itp.unibe.ch}
\affiliation[c]{Albert Einstein Center for Fundamental Physics, Institute for Theoretical Physics, University of Bern, 
Sidlerstrasse 5, CH-3012 Bern, Switzerland}

\author[d]{and Ville Vaskonen}
\emailAdd{ville.vaskonen@kcl.ac.uk}
\affiliation[d]{Physics Department, King's College London, London WC2R 2LS, United Kingdom}

\abstract{Making use of a dimensionally-reduced effective theory at high temperature, we perform a nonperturbative study of the electroweak phase transition in the Two Higgs Doublet model. We focus on two phenomenologically allowed points in the parameter space, carrying out dynamical lattice simulations to determine the equilibrium properties of the transition. We discuss the shortcomings of conventional perturbative approaches based on the resummed effective potential -- regarding the insufficient handling of infrared resummation but also the need to account for corrections beyond 1-loop order in the presence of large scalar couplings -- and demonstrate that greater accuracy can be achieved with perturbative methods within the effective theory. We find that in the presence of very large scalar couplings, strong phase transitions cannot be reliably studied with any of the methods. 
}

\maketitle

\newpage

\section{Introduction}

The great triumph of the Standard Model (SM) of particle physics was achieved in 2012 with the discovery of its last missing piece, the Higgs boson, by the ATLAS and CMS experiments at the Large Hadron Collider (LHC)~\cite{Aad:2012tfa,Chatrchyan:2012xdj}.
Although the properties of the observed Higgs boson so far agree with the SM predictions, it may be just one member of an extended Higgs sector.
Frameworks with non-minimal scalar sectors are amongst the best motivated beyond SM (BSM) scenarios, as they may provide solutions to many of the SM shortcomings, such as the origin of the observed baryon excess in the universe.

One of the most promising frameworks for producing this asymmetry is electroweak baryogenesis (EWBG), which produces the baryon excess during the electroweak phase transition (EWPT) at a temperature $T \sim 100\GeV$. Although the SM contains all the required ingredients for EWBG~\cite{Kuzmin:1985mm,Cohen:1990it,Turok:1990in}, it is unable to explain the observed  
baryon excess due to its insufficient amount of CP violation~\cite{Farrar:1993hn,Farrar:1993sp,Gavela:1994dt,Gavela:1994ds,Gavela:1993ts} and the lack of a first-order EWPT. Nonperturbative lattice studies in the SM have revealed that the Higgs boson is too heavy to lead to a large potential barrier between the symmetric and broken phases, and the electroweak symmetry breaking in the SM is thus a smooth crossover~\cite{Kajantie:1995kf,Kajantie:1996mn,Kajantie:1996qd,Csikor:1998eu,DOnofrio:2015gop}.

The study of EWBG is well-motivated in BSM models with extended Higgs sectors, which allow for new sources of CP violation 
and could provide a strongly first order phase transition. Among the simplest non-minimal Higgs frameworks are the Two Higgs Doublet Models (2HDMs), where the scalar sector is extended with one scalar doublet that has the same quantum numbers as the SM Higgs doublet. Both CP conserving and CP violating 2HDM frameworks have been studied in detail in the literature~\cite{Gunion:1989we,Gunion:2002zf,Haber:2006ue,Branco:2011iw,Keus:2015hva,Keus:2017ioh,Jung:2013hka,Brod:2013cka,Inoue:2014nva,Gaitan:2015aia,Chen:2015gaa,Ipek:2013iba}.

A common feature of BSM models with strongly first-order EWPT is that the relevant new fields can be light and hence dynamically active during the phase transition. This setup potentially leads to a multi-step transition with a tree-level potential barrier between the intermediate minimum and the final Higgs phase~\cite{Espinosa:2011ax,Espinosa:2011eu,Cline:2012hg,Patel:2012pi,Cline:2013gha,Blinov:2015sna}. Alternatively, radiative corrections from new fields strongly coupled to the Higgs boson can induce a large barrier between the origin and the Higgs phase and facilitate a strong single-step transition. In what follows, we will focus on the latter option and leave the discussion of multi-step phase transitions for future work. Strong phase transitions are interesting also because they can produce gravitational waves that may be observed in the near future~\cite{Caprini:2015zlo,Weir:2017wfa}. For both baryogenesis and gravitational-wave predictions, precise knowledge of the equilibrium properties of the EWPT is crucial.

In the context of the EWPT, variations of 2HDMs have been considered where the phase transition is analyzed using the perturbative effective potential~\cite{Cline:2011mm,Dorsch:2013wja,Dorsch:2014qja,Blinov:2015vma,Dorsch:2016nrg,Basler:2016obg,Basler:2017uxn,Chiang:2016vgf,Fuyuto:2017ewj,Laine:2017hdk}. Generically, in these works, a strongly first order EWPT is achieved through scalar couplings of $\mathcal{O}(1)$ or larger, which raises concerns of the performance of perturbation theory already at zero temperature. Additionally, at finite temperatures perturbative expansions suffer from severe infrared (IR) divergences in the presence of massless bosons~\cite{Linde:1980ts}. 
In particular, the symmetric Higgs phase is inherently nonperturbative and cannot be rigorously described by perturbative weak coupling methods, including the ring-improved perturbation expansions~\cite{Carrington:1991hz,Parwani:1991gq,Arnold:1992rz,Curtin:2016urg}.

The IR problem can be overcome with lattice Monte Carlo simulations. However, it is not known how to implement lattice fermions with non-Abelian chiral gauge couplings, rendering simulations of the full electroweak sector of the theory impossible~\cite{Luscher:2000hn}. Hence, the predominant approach for EWPT simulations is to make use of dimensionally-reduced effective theories (EFTs)~\cite{Ginsparg:1980ef,Appelquist:1981vg,Farakos:1994xh,Kajantie:1995dw} (see however~\cite{Csikor:1998eu}). In short, the EFT is obtained by integrating out non-zero Matsubara modes including all fermions, which have effective masses of order $2\pi T$ in the heat bath and decouple from the long-distance physics governing the phase transition. The EFT is then effectively three dimensional (hereafter 3d EFT), simplifying both perturbative and nonperturbative computations\footnote{In Refs.~\cite{Laine:1996ms,Losada:1996ju,Cline:1996cr,Cline:1997bm,Brauner:2016fla,Andersen:2017ika,Niemi:2018asa,Gould:2019qek}, it is discussed how existing lattice results in the EFT with one Higgs doublet can be applied to BSM theories using dimensional reduction, provided that the new degrees of freedom are sufficiently heavy. }.

In the paper at hand, we present a state-of-art study of the equilibrium dynamics of the EWPT, focusing on the CP-conserving 2HDM for simplicity. We carry out simulations with two dynamical doublets, overcoming the limitations of the previous analysis of Ref.~\cite{Andersen:2017ika} where only a limited region of the parameter space could be studied nonperturbatively. Since nonperturbative methods are very time consuming, we limit our analysis to two phenomenologically-motivated benchmark (BM) points where we carry out nonperturbative simulations, and perform a thorough comparison with conventional perturbative approaches. We find that for the moderately strong transitions studied here, the nonperturbative effects from IR physics are small in comparison to inaccuracies arising from bad convergence of perturbation theory due to the large scalar couplings, even at 2-loop level. As very strong transitions are generally associated with even larger couplings, our results suggests that in these cases perturbation theory fails to qualitatively describe the EWPT, and purely nonperturbative methods are then required.

The remainder of the paper is organized as follows. In section~\ref{sec:2hdm}
we review the scalar potential of the 2HDM and identify experimental and theoretical constraints applicable to our analysis. In section~\ref{sec:DR} we introduce the 3d EFT and discuss the basic ideas on which dimensional reduction is based. Our benchmark points for the lattice analysis are presented in section~\ref{sec:choosing_BMs}. Section~\ref{sec:lattice} is devoted to describing the lattice simulations, while in section~\ref{sec:pert_theory} we compare the perturbative and nonperturbative treatments of the model and justify the validity of our effective theory. Finally, in section~\ref{sec:conclusion} we draw our conclusions.

\section{The Two Higgs Doublet model}
\label{sec:2hdm}

We start by introducing the 2HDM. It consists of gauge and fermion sectors as in the SM, a scalar potential $V(\phi_1,\phi_2)$ for the two $\gr{SU(2)_L}$ scalar doublets with hypercharges $Y=1$ and their kinetic terms, as well as Yukawa interactions. We describe below the scalar potential and the structure of the Yukawa sector. 

\subsection{The scalar potential and the Yukawa sector}

In general, models with multiple Higgs doublets which can couple to fermions are at the risk of introducing Flavour Changing Neutral Currents (FCNCs) at tree level, which are tightly constrained by experiment. In the case of the 2HDM, imposing a $Z_2$ symmetry -- which can be softly-broken -- on the scalar potential and extending it to the fermion sector forbids these FCNCs. We therefore focus on a scalar potential of the form
\begin{eqnarray}
\label{eq:scalar_potential}
V(\phi_1,\phi_2) &=& \mu^2_{11} \he\phi_1 \phi_1 + \mu^2_{22} \he\phi_2 \phi_2 + \biggl[\mu^2_{12} (\phi_1^\dagger \phi_2) + \text{h.c.} \biggr] \nonumber \\
&+& \lambda_1 (\he\phi_1 \phi_1)^2 + \lambda_2 (\he\phi_2 \phi_2)^2 + \lambda_3 (\he\phi_1 \phi_1)(\he\phi_2 \phi_2)  + \lambda_4 (\he\phi_1 \phi_2)(\he\phi_2 \phi_1)  \nonumber \\
&+& \biggl[\frac{\lambda_{5}}{2}   (\phi_1^\dagger \phi_2)^2 + \text{h.c.} \biggr],
\end{eqnarray}
where $\mu^2_{12}$ causes a soft violation of the $Z_2$ symmetry, and explicitly $Z_2$-breaking terms of the form $(\phi^\dagger_1 \phi_2)(\phi^\dagger_1 \phi_1)$ and $(\phi^\dagger_2 \phi_1)(\phi^\dagger_2 \phi_2)$ have been discarded. In general, $\mu^2_{12}$ and $\lambda_5$ can be complex. We write the field composition of the two scalar doublets as
\begin{equation} 
\phi_1= \doublet{\phi^+_1}{\frac{1}{\sqrt{2}}\left(v_1+\rho_1+i \eta_1\right)} ,\quad 
\phi_2= \doublet{\phi^+_2}{\frac{1}{\sqrt{2}}\left(v_2+\rho_2+i \eta_2\right)} , 
\label{eq:fields}
\end{equation}
where the vacuum-expectation values (VEVs) $v_1$ and $v_2$ could in principle be complex.
 
The complex phases of $\lambda_5, \mu^2_{12}$ and the VEVs are connected to CP-violation in the scalar sector, relevant for baryon number violation during the EWPT. In the case a softly broken $Z_2$ symmetry, spontaneous CP violation can occur when $\Im(\lambda^*_5[\mu_{12}^2]^2)\neq 0$~\cite{Haber:2006ue,Haber:2015pua} and there exist no basis in which $\lambda_5$, $\mu_{12}^2$ and the VEVs are real. An exact $Z_2$ symmetry forbids the soft-breaking term $\mu_{12}^2$, which in turn leads to a real $\lambda_5$\footnote{This is known as {\sl rephasing invariance}~\cite{Ginzburg:2004vp}, which also removes the phases of the $v_i$'s in Eq.~(\ref{eq:fields}) by a redefinition of $\mu_{12}^2$ and $\lambda_5$ and renders the model CP conserving.}. However, CP-violating phases also contribute to the electric dipole moment
(EDM) and are hence heavily constrained by the strong bounds, in particular, on electron EDM from the ACME collaboration~\cite{Andreev:2018ayy}. As a result, baryogenesis does not seem feasible in a simple 2HDM framework. However, our goal is not to solve the full EWBG problem, but to study how accurately one can determine the main properties of the phase transition. Hence, in the paper at hand, we fix $\mu^2_{12}, \lambda_5$ as well as the VEVs to be real, and do not discuss CP violation further. In the absence of CP violation, the mass eigenstates $\{h, H, A, H^\pm\}$ can be written in terms of the fields in Eq.~(\ref{eq:fields}) and mixing angles $\alpha,\beta$ as 
\begin{equation}
\begin{aligned}
\label{eq:diagonalization}
h = -s_\alpha \rho_1 + c_\alpha \rho_2, \quad H = - c_\alpha \rho_1 - s_\alpha \rho_2, \\
H^{\pm} = -s_\beta \phi_1^{\pm} + c_\beta \phi_2^{\pm}, \quad A = -s_\beta \eta_1 + c_\beta \eta_2.
\end{aligned}
\end{equation} 
where $ s_\alpha = \sin\alpha,c_\alpha = \cos\alpha$, and the angle $\beta$ is related to the doublet VEVs by $\tan\beta = v_2/v_1$. 

The $Z_2$ charge assignment to fermions classifies four independent types of Yukawa interactions
which are known as Type-I, Type-II, Type-X and Type-Y\footnote{The Type-X and Type-Y 2HDMs 
are also referred to as the lepton-specific and flipped 2HDMs, respectively~\cite{Branco:2011iw}.}~\cite{Barger:1989fj,Grossman:1994jb,Akeroyd:1996he}. We shall study a Type-I 2HDM, where fermions are coupled only to the $\phi_2$ doublet. Consequently, constraints from flavor physics are less stringent than in Type-II, where down-type quarks are coupled to $\phi_1$ instead\footnote{In particular, the experimental bound on the mass of the charged scalar $H^{\pm}$ from $B$-physics in Type-II is $m_{H^{\pm}} \gtrsim 580 \GeV$~\cite{Misiak:2017bgg}, which is already so heavy compared to other mass scales in the theory that it could cause unnaturally large logarithmic corrections in the self energies.}. 

Finally, if the vacuum respects the $Z_2$ symmetry the $\phi_1$ doublet does not acquire a VEV, and the model is reduced to the Inert Doublet Model (IDM)~\cite{Ma:2006km,Barbieri:2006dq,LopezHonorez:2006gr}, for which a detailed EWPT analysis using a full 2-loop resummed effective potential has been performed in~\cite{Laine:2017hdk}.

\subsection{Input parameters for the numerical analysis}
\label{sec:inputs}

For our analysis, we renormalize the theory in the \MSbar scheme and treat $\tan\beta$ and $\alpha$ as input parameters, together with the softly $Z_2$-breaking parameter $\mu^2 \equiv -\mu^2_{12}$ and the physical pole masses $\{M_h, M_H, M_A, M_{H^\pm}\}$. These are related to the Lagrangian parameters via a 1-loop renormalization procedure described in detail in Appendix~\ref{sec:renormalization}. To summarize, the renormalized parameters are solved by requiring that the poles of loop-corrected propagators match the pole masses. Input parameters for the scalar sector are thus 
\begin{align}
\{M_h,\; M_H, \; M_A,\; M_{H^\pm},\; \tan\beta,\; \cos(\beta-\alpha),\; \mu^2,\; \FS\},
\end{align}
where the electromagnetic fine-structure constant $\FS$ essentially fixes the combination $v^2 \equiv v_1^2 + v_2^2$ at 1-loop level, and the gauge couplings $g, g'$ are fixed by loop corrections to the masses of $W^\pm, Z$ bosons. The scheme-dependent parameters $\tan\beta, \alpha$ and $\mu^2$ are input at a fixed \MSbar scale. However, we will also discuss the EWPT in the presence of an inert $\phi_1$ ($\mu^2=v_1=0$). In this case, we input the \MSbar couplings $\lambda_1$ and $\lambda_{345}/2 \equiv (\lambda_3+\lambda_4+\lambda_5)/2$, corresponding to self-interaction and portal coupling of the dark matter candidate $H$~\cite{Ma:2006km,Barbieri:2006dq,LopezHonorez:2006gr}, instead of the mixing angles. In what follows, we identify $h$ as the observed scalar with $M_h = 125.09 \GeV$.

The quantity $\cos(\beta-\alpha)$ is important for 2HDM phenomenology as it controls the interaction strengths of the CP-even scalars $h,H$ to electroweak gauge bosons. The case $\cos(\beta-\alpha) = 0$ corresponds to the alignment limit where $h$ couples to SM particles exactly as the physical Higgs in the SM. In practice, 2HDMs are driven to the alignment limit by constraints from collider experiments~\cite{Agashe:2014kda,Khachatryan:2014iha,Haller:2018nnx}. Additional particles may introduce important radiative corrections to gauge boson propagators. Furthermore, electroweak precision measurements of the oblique parameters~\cite{Peskin:1991sw,Peskin:1990zt,Grimus:2008nb,Haber:2010bw} are satisfied when the charged scalar ${H^\pm}$ is close in mass with either $H$ or $A$~\cite{Gerard:2007kn,Haber:2010bw,Funk:2011ad}. In our analysis, we consider the $M_{H^\pm} = M_A$ case.

In the phase transition analysis, we take into account the top Yukawa coupling, $y_t$, to $\phi_2$ and neglect the Yukawa couplings of other fermions due to their subdominant contribution. As a result, our EWPT analysis is valid for 2HDMs of Type-I\footnote{In the Type II model, the down-type Yukawa couplings obtain a large enhancement when $\tan\beta \gtrsim 1$ and cannot be neglected.}. It is worth mentioning that the light Yukawa couplings are included in our renormalization procedure, where we assume Type-I Yukawa couplings, but have verified that their numerical effect on the self energies is negligible.

\section{High-temperature dimensional reduction}
\label{sec:DR}

The concept of dimensional reduction in thermal field theory is based on the observation that a quantum system in a heat bath possesses a natural scale hierarchy.
In the Matsubara formalism, this hierarchy can be made explicit by Fourier expanding the fields with respect to the imaginary time. The theory can then be described in terms of 3d Fourier modes, where the modes with a non-zero Fourier frequency obtain a mass correction of order $2\pi T$ in the propagators. These non-zero Matsubara modes can be integrated out perturbatively in an IR safe manner, and the resulting 3d EFT for the IR-sensitive Matsubara zero modes can then be studied nonperturbatively on the lattice. In general, the EFT is purely bosonic due to the lack of fermionic zero-modes, and numerical simulations are straightforward.

\subsection{Three-dimensional effective theory for the 2HDM}
\label{sec:3d-EFT}

The dimensionally-reduced EFT for the CP-conserving 2HDM has been derived in Ref.~\cite{Gorda:2018hvi}, extending the previous derivations of Refs.~\cite{Losada:1996ju,Andersen:1998br}, and has the form
\begin{align}
\label{eq:eff_theory}
\mathcal{L}^\text{(3d)} =& \frac14(F_{rs})^2_\text{3d} + (D_r\phi)_\text{3d}^\dagger (D_r\phi)_\text{3d} \nonumber \\
+& \bar{\mu}^2_{11} (\he\phi_1 \phi_1)_\text{3d} + \bar{\mu}^2_{22} (\he\phi_2 \phi_2)_\text{3d} + \bar{\mu}^2_{12} \Big((\he\phi_1 \phi_2)_\text{3d} + (\he\phi_2 \phi_1)_\text{3d}\Big) \nonumber \\
+& \bar{\lambda}_{1} (\he\phi_1 \phi_1)_\text{3d}^2 + \bar{\lambda}_{2} (\he\phi_2 \phi_2)_\text{3d}^2 + \bar{\lambda}_3 (\he\phi_1 \phi_1)_\text{3d}(\he\phi_2 \phi_2)_\text{3d}  + \bar{\lambda}_4 (\he\phi_1 \phi_2)_\text{3d}(\he\phi_2 \phi_1)_\text{3d}  \nonumber \\
+& \frac{\bar{\lambda}_5}{2} \left( (\he\phi_1 \phi_2)_\text{3d}^2  + (\he\phi_2 \phi_1)_\text{3d}^2\right),
\end{align}
where the (three-dimensional) field-strength tensor $F_{rs}$ and the covariant derivative $D_r$ are understood to contain both the $\gr{SU(2)}$ and $\gr{U(1)}$ gauge fields with the gauge couplings denoted by $\bar{g}$ and $\bar{g}'$. We have suppressed the gluon and ghost terms which are irrelevant for the EWPT, as they do not directly couple to the scalars\footnote{Ghosts do, however, appear in the perturbative calculation of section~\ref{sec:3d-Veff}}. Having effectively integrated out the temporal direction, this theory is defined in a three-dimensional Euclidean space and contains only the zero Matsubara frequency modes of the original fields. We denote the fields with the subscript $\text{3d}$ to emphasize this fact. Furthermore, we absorb the factor $1/T$ multiplying the action into the field definitions, so that the fields have the dimension $\text{GeV}^{1/2}$ and all couplings have a positive mass dimension. Higher-order operators, such as $(\phi^\dagger_1\phi_1)_{\text{3d}}^3$, have been dropped from Eq.~(\ref{eq:eff_theory}); we will discuss these operators in section~\ref{sec:DR-validity}.

This EFT is constructed perturbatively by matching the Green's functions in both theories at $\mathcal{O}(\lambda_i^2)$ accuracy in the scalar couplings and $\mathcal{O}(g^4), \mathcal{O}(y^4_t)$ in the gauge and top Yukawa couplings. This corresponds to a 1-loop matching of four-point functions and 2-loop matching of the scalar two-point functions. We use high-$T$ expansion in computation of the sum-integrals, which leads to additional contributions of order $\mathcal{O}(\mu^2_i \lambda_j), \mathcal{O}(\mu^2_i g^2), \mathcal{O}(\mu^2_i y^2_t)$ that are also contained in the matching relations. Let us note that the construction of the theory in Eq.~(\ref{eq:eff_theory}) also involves integrating out the temporal components of the gauge fields, which generate effective masses of order $gT$ due to Debye screening. This results in a small correction to the EFT parameters. A detailed derivation has been presented in Ref.~\cite{Gorda:2018hvi}; in particular, see section 3.3 there for explicit matching relations for the couplings of the effective theory.

We emphasize that the 3d EFT approach is useful not only for lattice simulations, but also as a way of organizing perturbation theory. The reason is that thermal resummations beyond 1-loop order are automatically implemented in the parameter matching. Indeed, the renormalized masses $\bar{\mu}^2_{11}, \bar{\mu}^2_{22}$ are just the screened masses evaluated at 2-loop level~\cite{Braaten:1995cm}. Loop calculations are also simplified, as the 3d EFT is purely spatial and contains one mass scale less than the full theory (the temperature), and is furthermore super-renormalizable.

\subsection{Lattice formulation of the effective theory}

Our discretized action in three dimensions, corresponding to the 3d EFT in Eq.~(\ref{eq:eff_theory}), reads 
\begin{align}
\label{eq:lattice_action}
S_\text{lat} =& \beta_G \sum_\vec{x} \sum_{i<j} (1 - \frac12 \text{Tr} \, P_{ij}(x)) 
\nonumber \\
-& 2a\sum_\vec{x} \sum_{i} \RE\Big[ \Phi_1^\dagger(x) U_i(x) \Phi_1(x+i) + \Phi_2^\dagger(x) U_i(x) \Phi_2(x+i)\Big] \nonumber \\
& + \sum_\vec{x} a^3 \Big[\left(\frac{6}{a^2} + m^2_{11}\right) \Phi_1^\dagger \Phi_1 + \left(\frac{6}{a^2} + m^2_{22}\right) \Phi_2^\dagger \Phi_2 + m^2_{12}  \left(\Phi_1^\dagger \Phi_2 + \Phi_2^\dagger \Phi_1\right) \nonumber \\
& + \bar{\lambda}_1 (\Phi_1^\dagger \Phi_1 )^2 + \bar{\lambda}_2 (\Phi_2^\dagger \Phi_2 )^2 + \bar{\lambda}_3 (\Phi_1^\dagger \Phi_1 )(\Phi_2^\dagger \Phi_2 ) + \bar{\lambda}_4 (\Phi_1^\dagger \Phi_2 )(\Phi_2^\dagger \Phi_1 )
\nonumber \\
& +  \frac{\bar{\lambda}_5}{2} \left( (\Phi_1^\dagger \Phi_2 )^2 + (\Phi_2^\dagger \Phi_1 )^2 \right)\Big], 
\end{align}
where $a$ is the lattice spacing and the dimensionless constant $\beta_G$ is given by
\begin{align}
\beta_G = \frac{4}{a \bar{g}^2}.
\end{align}
In Eq.~(\ref{eq:lattice_action}), $U_i(x)$ are the $\gr{SU(2)}$ gauge links and $P_{ij}$ is the standard Wilson plaquette. Following previous lattice studies of the EWPT~\cite{Farakos:1994xh,Kajantie:1995kf,Laine:1998qk,Laine:1998vn,Laine:2000rm}, we have dropped the $\gr{U(1)}$ gauge field from the lattice action as its effect on the dynamics of the transition is small~\cite{Kajantie:1996qd}.  

The masses $m^2_{11}, m^2_{22}, m^2_{12}$ and fields $\Phi_1, \Phi_2$ in the lattice action are related to the corresponding continuum quantities $\bar{\mu}^2_{11}, \bar{\mu}^2_{22}, \bar{\mu}^2_{12}$ and $(\phi_1)_\text{3d}, (\phi_2)_\text{3d}$ in the \MSbar scheme by relations that can be found in Ref.~\cite{Laine:2000rm} (see Appendix B therein). We emphasize that due to the super-renormalizability of the effective 3d theory, Eq.~(\ref{eq:eff_theory}), these lattice-continuum relations are exact and not susceptible to perturbative errors~\cite{Laine:1995np,Laine:1997dy}. Couplings in the 3d theory do not run, so their lattice-continuum relations are trivial. For the actual simulations, we find it convenient to make the fields and couplings dimensionless by scaling them with appropriate factors of $T$ as in Ref.~\cite{Laine:2000rm}.

\section{Choosing the benchmark points}
\label{sec:choosing_BMs}

Because of the computational effort required for lattice simulations, it is not possible to perform nonperturbative scans over the whole parameter space allowed by theoretical and experimental constraints. We therefore need to focus our analysis on a couple of points from which one can hope to draw more general conclusions about the performance of perturbation theory. Let us reiterate that in order to generate a potential barrier for a strong single-step EWPT, some of the Higgs sector couplings will necessarily have to be large. Unfortunately, in many strong-EWPT scenarios present in the literature, some of these couplings are so large that the convergence of perturbation theory is at best marginal (cf. section~\ref{sec:DR-validity}), and care needs to be taken when constructing the 3d EFT perturbatively.

In order to guarantee the accuracy of our 3d EFT, the couplings should be kept small enough so that loop corrections from the heavy Matsubara modes remain under control. Furthermore, the thermal scale hierarchy should be respected, so that all scalar degrees of freedom have to be lighter than $2\pi T$ in both phases near the critical temperature, where $\phi_c \sim T_c$ for strong transitions.

\subsection{2HDM scenarios to be studied on the lattice}
\label{sec:benchmarks}

With the above considerations in mind, we have chosen two phenomenologically-viable BM scenarios -- described in Table~\ref{table:input-masses} -- where we expect the 3d EFT to accurately describe the EWPT. In addition to verifying boundedness from below of the scalar potential at 1-loop level (see section~\ref{sec:MSbar-Veff}), we have checked that tree-level unitarity constraints~\cite{Kanemura:1993hm,Akeroyd:2000wc} are satisfied and that the largest coupling $\lambda_3$ stays below $2\pi$
at scales relevant for the EWPT\footnote{The perturbativity bound $\lambda_i < 2\pi$ is motivated by Refs.~\cite{Nierste:1995zx,Braathen:2017izn,Krauss:2017xpj}, where the breakdown of perturbation theory is demonstrated for couplings much smaller than the naive upper bound of $4\pi$. In practice, the magnitude of our largest coupling $\lambda_3$ in BM2 is roughly $\lambda_3 \sim 4$ at the scale of EWPT dynamics, with the other couplings being substantially smaller. } (c.f. section~\ref{sec:procedure}). Although some of the vacuum masses in both benchmarks are heavy compared to the electroweak scale, the high-$T$ expansion used in dimensional reduction can still be expected to converge well as the fields are lighter near the phase transition~\cite{Laine:2000kv,Laine:2017hdk}.

\begin{table}[h!]
\centering
  \begin{tabular}{| l l l l l l l l |}
	\hline
	\multirow{2}{*}{BM1} & 
    $M_H$ & $M_A$ & $M_{H^\pm}$ & $\mu$ & $\lambda_{345}/2$ & $\lambda_1$ & $\Lambda_0$ \\
     & $66$\,GeV & $300$\,GeV  & $300$\,GeV & $0$\,GeV & $1.07\times 10^{-2}$ & $0.01$ & $91$\,GeV \\
\hline
	\multirow{2}{*}{BM2} &
	$M_H$ & $M_A$ & $M_{H^\pm}$ & $\mu$ & $\cos(\beta-\alpha)$ & $\tan\beta$ & $\Lambda_0$\\
     & $150$\,GeV & $350$\,GeV  & $350$\,GeV & $80$\,GeV & $-0.02$ & $2.75$ & $265.018$\,GeV\\
\hline
\end{tabular}
  \caption{Input parameters for our benchmark points. In BM1, the combination $\lambda_{345} \equiv \lambda_3+\lambda_4+\lambda_5$ corresponds to a dark matter portal coupling in the IDM~\cite{Blinov:2015vma}, and $\mu=\sqrt{-\mu^2_{12}}$ represents the soft $Z_2$-breaking parameter. Masses are assumed to be the physical pole masses, while the remaining parameters are input directly in the \MSbar scheme at the initial renormalization scale $\Lambda_0$.}
  \label{table:input-masses}
\end{table}

BM1 is specific to the IDM and has been studied perturbatively in Refs.~\cite{Blinov:2015vma,Laine:2017hdk}. Our main motivation for studying this particular point on the lattice is to produce a quantitative comparison with the resummed 2-loop result of Ref.~\cite{Laine:2017hdk}, where the 2-loop corrections to the effective potential were found to make the transition considerably weaker relative to a 1-loop calculation. To make this comparison, we have modified our renormalization procedure to match that of Ref.~\cite{Laine:2017hdk}; In particular, we neglect the $\gr{U(1)}$ sector by setting $g'=0$ in BM1, but have numerically verified that its inclusion in the 1-loop calculation has a negligible effect on the renormalized parameters listed in Table~\ref{table:input-params}. On the phenomenological side, BM1 provides a dark matter candidate $H$, which can constitute a fraction of the observed dark matter relic density~\cite{Blinov:2015vma}.

BM2, in the softly $Z_2$-breaking 2HDM, lies in the mass-hierarchy region where earlier studies based on the 1-loop effective potential report strong first-order phase transitions~\cite{Dorsch:2013wja,Dorsch:2017nza,Basler:2016obg}. Our BM2 approaches the parameter-space points in which Refs.~\cite{Dorsch:2016nrg,Caprini:2015zlo} predict gravitational-wave signatures in the sensitivity range of LISA. However, BM2 represents a more conservative EWPT scenario than what is shown in {\em e.g.}~\cite{Dorsch:2016nrg} with a large $\lambda_3$, but where perturbation theory still converges reasonably well (modulo the usual IR problems), while also providing a moderately strong transition. 

Phenomenologically, BM2 is motivated by possible collider signatures in the following processes, away from the alignment limit and in the small $\tan\beta$ region: the ratio of decay rates of $h$ (the SM-like Higgs boson) to those of the $h^{SM}$ (the Higgs boson in the SM) in $b\bar{b}, \tau^+ \tau^-, gg$ channels, as well as the signal strength of the SM-like Higgs boson in the $\tau^+ \tau^-$ decay mode\footnote{The signal strength is defined as $\mu_{\tau^+ \tau^-}=\frac{\sigma(gg\to h)}{\sigma(gg \to h^{SM})}\times \frac{BR(h\to \tau^+ \tau^-)}{BR(h^{SM} \to \tau^+ \tau^-)}$.}. Complementary to those, the $hW^+$ and $HW^+$ or $AW^+$ decays of the heavy charged Higgs produce interesting experimental signatures for this BM point \cite{Keus:2015hva}. Finally, the $A \rightarrow Z h$ channel acts as an extra probe of this BM point~\cite{Dorsch:2014qja,CMS:1900sig}.

Experimental constraints on 2HDMs come from gauge bosons width~\cite{Agashe:2014kda}, direct searches for charged scalars and lifetime of charged scalars~\cite{Pierce:2007ut}, Higgs total decay width~\cite{Khachatryan:2014iha}, Higgs invisible branching ratios and Higgs to $\gamma\gamma$ signal strength~\cite{TheATLASandCMSCollaborations:2015bln,CMS:2015kwa}, $A \to Z h$ searches~\cite{Aad:2015wra}, direct searches for extra Higgs bosons at the LHC~\cite{Aad:2014vgg} and flavour constraints~\cite{Mahmoudi:2009zx}. We have verified that our benchmarks satisfy all current experimental bounds arising from these sources.

In Table~\ref{table:input-params}, we list the renormalized parameters, obtained from the input parameters, as described in Appendix~\ref{sec:renormalization}. We have chosen different input \MSbar scales for BM1 and BM2 points. In BM1, the parameters in Table~\ref{table:input-params} are solved from the loop-corrected pole-mass conditions at scale $M_Z \approx 91$ GeV, in accordance with Ref.~\cite{Laine:2017hdk}. In BM2, however, we choose the initial scale to be the average of the pole masses, $\Lambda_0 = (M_h + M_H + M_A + 2M_{H^\pm})/5$, in order to reduce the size of logarithmic corrections in the self energies. This choice for the input scale is justified by the numerical analysis of Ref.~\cite{Altenkamp:2017ldc}, where the scale dependence of different renormalization prescriptions is discussed.

\begin{table}[H]
\centering
	\begin{tabular}{l l l}
	& BM1 & BM2 \\ 
	\hline
	$\mu^2_{11}$/GeV$^2$ & $842$ & $12942$ \\
	$\mu^2_{22}$/GeV$^2$ & $-6669$ & $-2751$ \\
	$\mu^2_{12}$/GeV$^2$ & $0$ & $-6400$ \\
	$\lambda_1$ & $0.010$ & $0.300$ \\
	$\lambda_2$ & $0.0670$ & $0.0925$ \\
	$\lambda_3$ & $2.757$ & $3.675$ \\
	$\lambda_4$ & $-1.368$ & $-1.780$ \\
	$\lambda_5$ & $-1.368$ & $-1.792$ \\
	$g^2$ & $0.425$ & $0.418$ \\
	${g'}^2$ & $0$ & $0.130$ \\
	$g^2_s$ & $1.489$ & $1.489$ \\
	$y^2_t$ & $0.971$ & $0.998$ \\
	\end{tabular}
  \caption{Renormalized parameters corresponding to the input parameters in Table~\ref{table:input-masses}. The recipe for obtaining these is described in the main text and in Appendix~\ref{sec:renormalization}. In BM1, we have set the $\gr{U(1)}$ coupling to zero for the sake of comparison with the results of Ref.~\cite{Laine:2017hdk}. The $\gr{SU(3)}$ coupling $g_s$ is fixed at tree level.}
  \label{table:input-params}
\end{table}

\section{Lattice simulations}
\label{sec:lattice}

The discretized theory, Eq.~(\ref{eq:lattice_action}), is studied nonperturbatively by evaluating expectation values of quantum operators using Monte Carlo integration. Our simulations are performed with the same Monte Carlo code that was used in Refs.~\cite{Kajantie:1995kf,Laine:1998vn,Laine:2000rm}, and the practical procedure is similar to that of Ref.~\cite{Laine:2000rm} where the Minimally Supersymmetric Standard Model (MSSM) was studied on the lattice. We emphasize that since gauge fixing is not needed on the lattice, the results obtained here are manifestly gauge invariant.

\subsection{Obtaining lattice parameters}
\label{sec:procedure}

Starting from a set of given input parameters, our analysis proceeds as follows. In order to account for logarithmic corrections of thermal origin, we first run the parameters in Table~\ref{table:input-params} to the scale 
\begin{align}
\label{eq:DR-scale}
\Lambda_\text{DR} = 4\pi e^\gamma T \approx 7.055 T,
\end{align}
with $\gamma$ being the Euler-Mascheroni constant, using 1-loop $\beta$ functions found {\em e.g.} in Refs.~\cite{Branco:2011iw,Gorda:2018hvi}. Eq.~(\ref{eq:DR-scale}) has the physical interpretation of corresponding to the average momentum of integration over the non-zero Matsubara modes~\cite{Farakos:1994kx}. We then apply the matching relations of Ref.~\cite{Gorda:2018hvi} to obtain the parameters of the effective theory, Eq.~(\ref{eq:eff_theory}), as functions of the temperature, which are converted to lattice parameters using the relations provided in Ref.~\cite{Laine:2000rm}. 

When the RG scale is chosen as in Eq.~(\ref{eq:DR-scale}), all logarithmic corrections in the matching relations vanish, apart from those corresponding to the (exact) RG running of the masses within the 3d theory~\cite{Farakos:1994kx,Gorda:2018hvi}. RG scale in the effective theory is separate from that of the full theory and we fix it as $\Lambda_{\text{3d}} = T$, although the lattice results are insensitive to this choice as RG running is exactly contained in the lattice-continuum relations.

\subsection{Finding the transition point on the lattice}

From the simulations, we obtain the gauge-invariant expectation values of the operators in Eq.~(\ref{eq:lattice_action}) for fixed volume $V$ and $\beta_G$ which sets the lattice spacing. At the critical point, the metastability of the phases is so strong that normal simulation methods do not efficiently tunnel between the phases. Hence, we apply multicanonical simulations~\cite{Berg:1991cf} to overcome the potential barrier suppressing mixed-phase configurations. In both of our BM points, the $\Phi_2$ field (in lattice discretization) dominates the phase transition dynamics, while the other doublet $\Phi_1$ is so heavy that its condensate, $\langle \Phi^\dagger_1\Phi_1 \rangle$, changes only slightly at the transition point. We can therefore treat the expectation value $\langle \Phi^\dagger_2\Phi_2 \rangle$ as an effective order parameter that determines the transition point. 

\begin{figure}[H]
\centering
     \subfloat[BM1, $\beta_G = 16$]{\includegraphics[scale=0.3725]{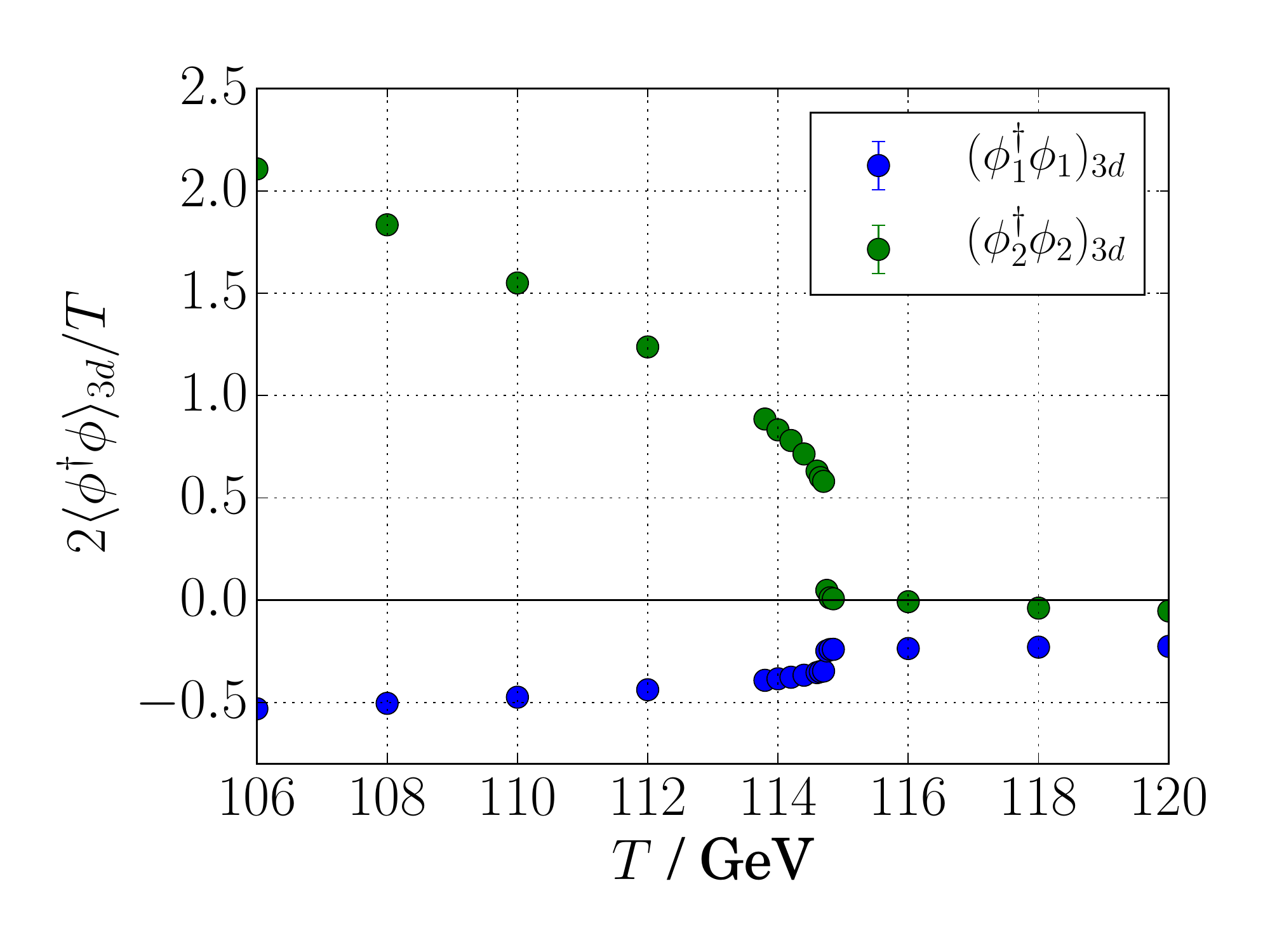}\label{}}
		\kern-0.5em
     \subfloat[BM2, $\beta_G = 28$]{\includegraphics[scale=0.3725]{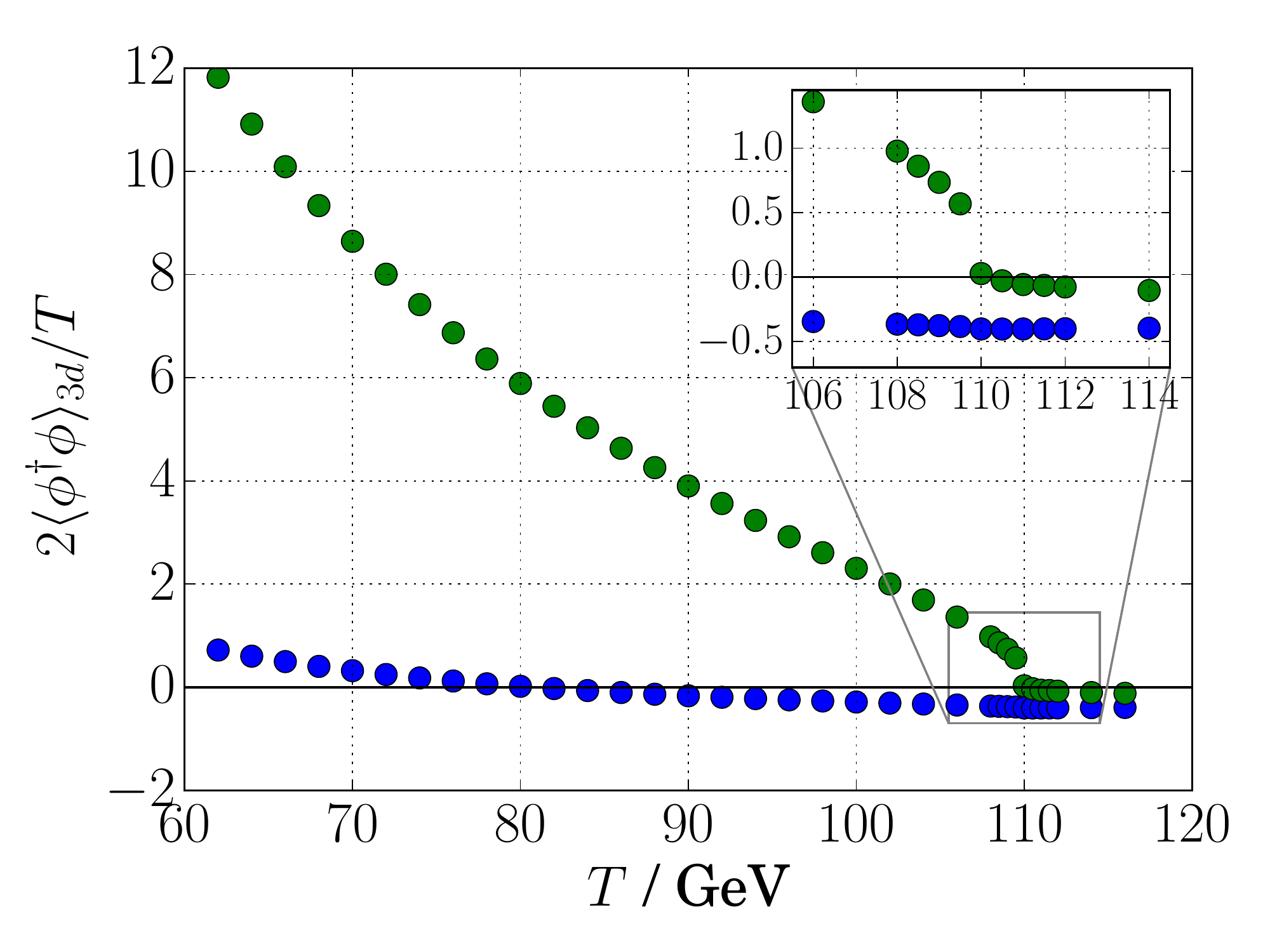}\label{}}
     \caption{Gauge-invariant condensates of the two doublets as measured on the lattice with fixed volume and $\beta_G$, converted to \MSbar quantities using the relations in Ref.~\cite{Laine:1997dy}. In both cases, the doublet $\phi_1$ is heavy and almost inert at the phase transition. Instead, its fluctuations become more constrained as $\langle \phi^\dagger_2 \phi_2\rangle$ changes due to the $\lambda_3$ term in the action. In BM2, $\langle \phi^\dagger_1 \phi_1\rangle$ increases smoothly towards its $T=0$ value as governed by our choice of $\tan\beta$. }
	 \label{fig:condensates_lat}
\end{figure}

The composite operators are divergent in the ultraviolet (UV) and can hence obtain negative values when renormalization is applied. The behavior of the condensates is plotted in Fig.~\ref{fig:condensates_lat}, where we have converted the lattice fields $\Phi^\dagger_i\Phi_i$ to the respective quantities in the \MSbar scheme, $(\phi^\dagger_i\phi_i)_\text{3d}$, by substracting the lattice divergence~\cite{Laine:1997dy}. We shall drop the field subscripts in the following discussion. In BM1, the change in the $\phi^\dagger_1 \phi_1$ condensate is a result of field fluctuations becoming more constrained due to the $\phi_2$ field obtaining a VEV. In BM2, on the other hand, $\phi_1$ will also develop a VEV due to the requirement that $\tan\beta$ is non-zero in the $T=0$ vacuum. This change is compensated by decreased fluctuations around the new minimum, and as a result, the condensate $\langle \phi^\dagger_1\phi_1 \rangle$ remains in practice constant in the transition.   

\begin{figure}[h]
\centering
\includegraphics[scale=0.6]{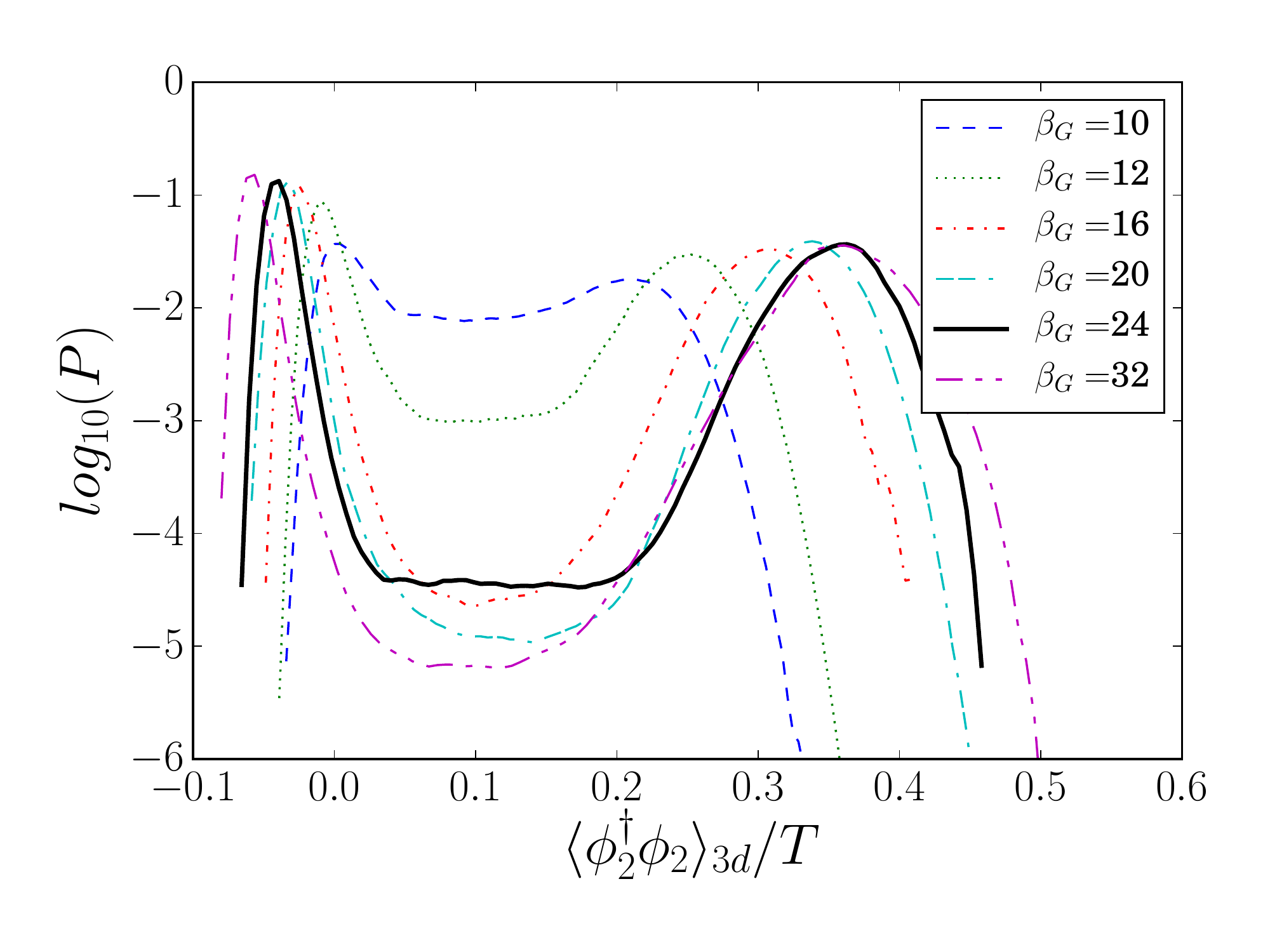}
\caption{Unnormalized probability distributions of the expectation value $\langle \phi^\dagger_2 \phi_2 \rangle_\text{3d}$ in BM1, measured on the lattice with varying $\beta_G$ and converted to the corresponding continuum quantity. The histograms have been obtained by reweighting in $T$ and minimizing the integrated probability difference between the peaks. As $\beta_G$ increases, the phases become more separated and thus the transition grows stronger. Configurations between the two phases are suppressed exponentially by the free energy carried by the phase interface, with the suppression increasing with the area of the interface.} 
\label{fig:BM1_hgram}
\end{figure}

In a first-order phase transition, the probability distribution of $\langle \phi^\dagger_2\phi_2 \rangle$ has a two-peak structure such as the one shown in Fig.~\ref{fig:BM1_hgram}, with the peaks corresponding to the symmetric and broken phases. The probability of field configurations between the peaks is strongly suppressed, and the separation of the peaks corresponds to a potential barrier that the system has to overcome in order for the phase transition to occur. At the critical temperature $T_c$, the probability of finding the system in either phase is equal. Hence, our criterion for finding $T_c$ is that the integrated probability under the peaks in the histogram is the same. In practice, the simulations are carried out at a temperature close to $T_c$, and the precise critical temperature is then conveniently found by reweighting~\cite{Ferrenberg:1988yz} the multicanonical distribution to a temperature that minimizes the integrated probability difference of the two phases.

\subsection{Determining physical observables from the simulations}
\label{sec:lattice_results}

In both BM points, properties of the transition depend strongly on the lattice spacing, or $\beta_G$, as well as on the volume to a lesser extent. Continuum results for the equilibrium quantities characterizing the phase transition are obtained by first extrapolating to an infinite volume, and taking the lattice spacing to zero afterwards (corresponding to $1/\beta_G \rightarrow 0$). The simulated values of $V$ and $\beta_G$ are listed in Table~\ref{table:simulations}. Each simulation has been weighted by an appropriate multicanonical weight function and consists of $1.5 \times 10^6 - 2.5 \times 10^6$ measurements, depending on the volume. Cylindrical lattices have been used for a precise measurement of the interface tension (see below).

\begin{figure}[H]
\centering
\subfloat[BM1]{\scalebox{1}
{\begin{tabular}{|l| l l l l|} 
	\hline
    $\beta_G$ & \multicolumn{4}{c|}{Volumes, $L_x \times L_y \times L_z$} \\
	\hline
	$10$ & $18^2 \times 72$ & $20^2 \times 80$ & $24^2 \times 96$ & \\
	$12$ & $20^2 \times 96$ & $24^2 \times 96$ & $28^2 \times 120$ & \\
	$14$ & $28^2 \times 84$ & $28^2 \times 140$ &  & \\ 
	$16$ & $24^2 \times 96$ & $32^2 \times 120$ & $32^2 \times 162$ & \\ 
	& $38^2 \times 162$ & & &  \\
	$20$ & $24^2 \times 112$ & $32^2 \times 132$ & $38^2 \times 156$ & \\
	$24$ & $34^2 \times 156$ & $42^2 \times 172$ & $42^2 \times 200$ & \\
	$32$ & $42^2 \times 200$ & $48^2 \times 192$ & $54^2 \times 216$ & \\
	\hline
  \end{tabular}
}} \quad
\subfloat[BM2]{\scalebox{1}
{\begin{tabular}{|l| l l l|} 
	\hline
    $\beta_G$ & \multicolumn{3}{c|}{Volumes, $L_x \times L_y \times L_z$} \\
	\hline
	$20$ & $32^2 \times 132$ & $38^2 \times 156$ & $42^2 \times 168$  \\
	$24$ & $34^2 \times 156$ & $42^2 \times 172$ & $48^2 \times 182$  \\
	$28$ & $42^2 \times 168$ & $48^2 \times 192$ & $54^2 \times 200$  \\
	$32$ & $48^2 \times 192$ & $54^2 \times 216$ & $58^2 \times 240$  \\
	\hline
  \end{tabular}
}}
\caption{Lattice volumes and values of $\beta_G$ used in simulations. The volumes are given in units of $a^3$. We use lattices cylindrical in the $z$ direction, with the remaining two directions having equal length. In BM2, some scalar degrees of freedom are so heavy that it is necessary to use small lattice spacings (large $\beta_G$) to fully capture their effect on the phase transition. Simulations with $\beta_G=14$ in BM1 are only used in the continuum extrapolation of $T_c$. }
\label{table:simulations}
\end{figure}

Our results are collected in Table~\ref{table:lattice_results} along with statistical errors, obtained with jackknife sampling, related to the continuum extrapolations. The behavior of the extrapolations is qualitatively similar to the MSSM case of Ref.~\cite{Laine:2000rm}, however in our simulations, the latent heat and order parameter discontinuity contain substantial dependence on the lattice volume.

\begin{table}[H]
\centering
  \begin{tabular}{l|l|l|l|l}
    & $T_c$/GeV & $L/T^4_c$ & $\Delta \phi/T$ & $\sigma/T^3_c$  \\
\hline
BM1 & $116.4021 \pm 0.0047$ & $0.603 \pm 0.023$ & $1.075 \pm 0.021$ & $0.0270 \pm 0.0013$  \\
BM2 & $112.4540 \pm 0.0145$ & $0.807 \pm 0.051 $ & $1.087 \pm 0.034$ & $0.0204 \pm 0.0045$
  \end{tabular}
  \caption{Nonperturbatively determined critical temperature, latent heat, order parameter discontinuity and surface tension of the phase transition. The errors shown here are statistical errors of least square polynomial fitting.}
  \label{table:lattice_results}
\end{table}

\paragraph{Critical temperature:} 

\begin{figure}[H]
\centering
     \subfloat[BM1]{\includegraphics[scale=0.3725]{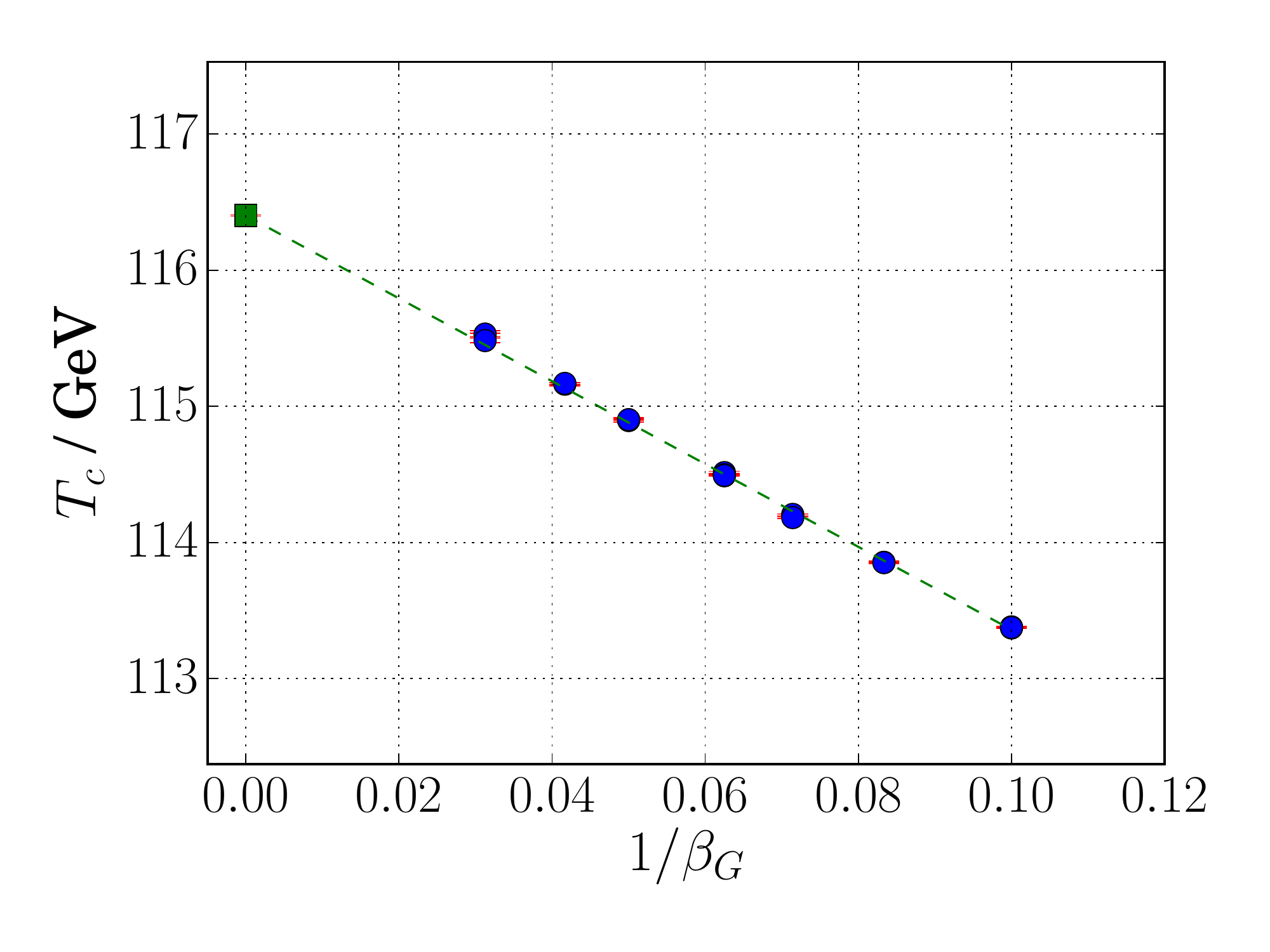}\label{}}
		\kern-0.5em
     \subfloat[BM2]{\includegraphics[scale=0.3725]{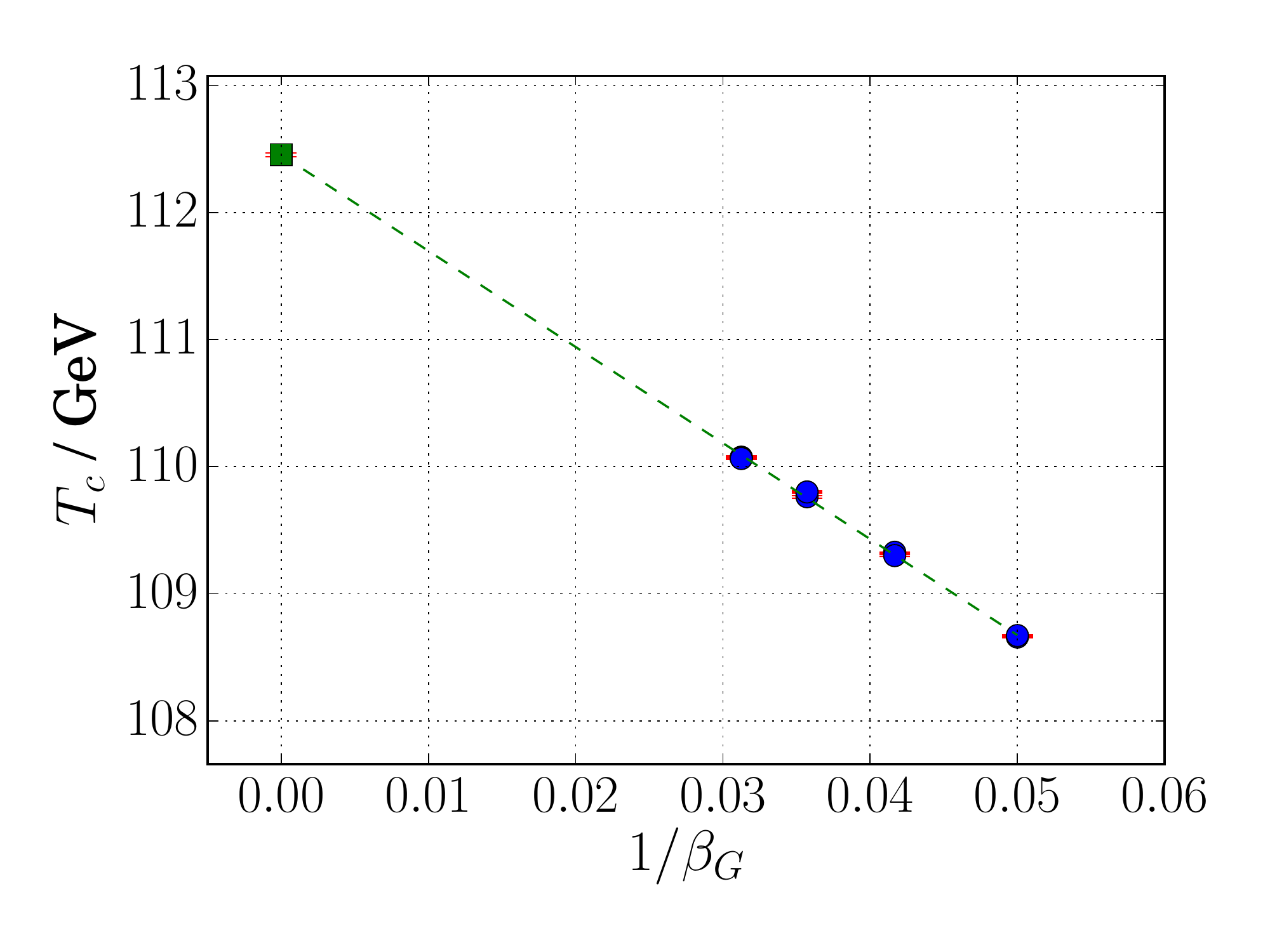}\label{}}
     \caption{Continuum extrapolation of the critical temperature. All volumes are plotted here and are indistinguishable from each other at the same $\beta_G$.}
	 \label{fig:Tc_lat}
\end{figure}

For individual simulations with fixed volume and $\beta_G$, the critical temperature is determined from equal-weight histograms as described above. $T_c$ in both BM points is insensitive to the volume, and the dependence on $1/\beta_G \propto a$ is linear over the entire range of the lattice spacings used. Extrapolations to continuum are shown in Fig.~\ref{fig:Tc_lat}. Statistical errors from the Monte Carlo simulations -- as well as those from reweighting -- are small for the critical temperature. 

\paragraph{Discontinuity in the order parameter:} 

\begin{figure}[h] 
	\centering
     \subfloat[BM1]{\includegraphics[scale=0.3725]{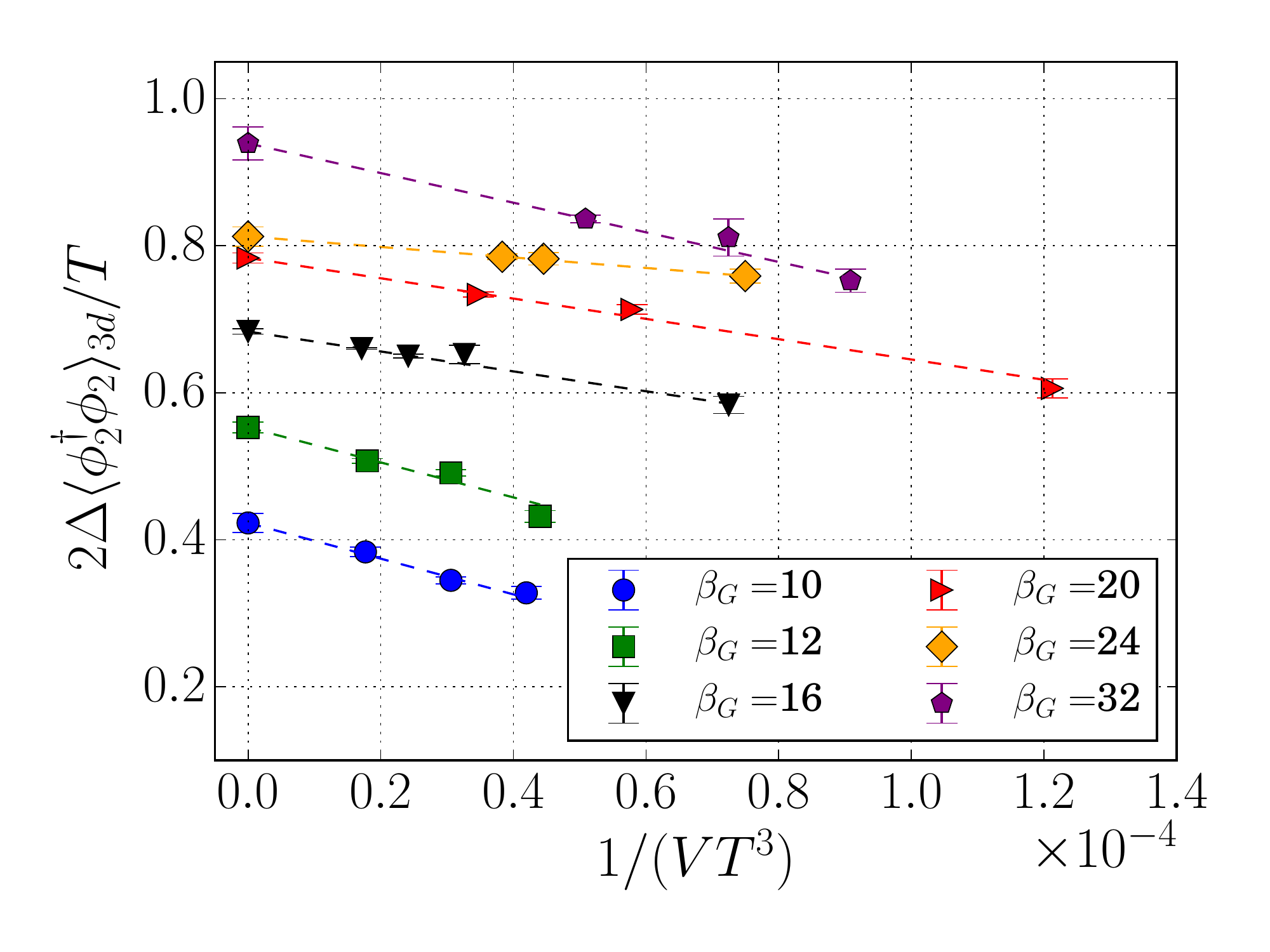}\label{}}
		\kern-0.5em
     \subfloat[BM2]{\includegraphics[scale=0.3725]{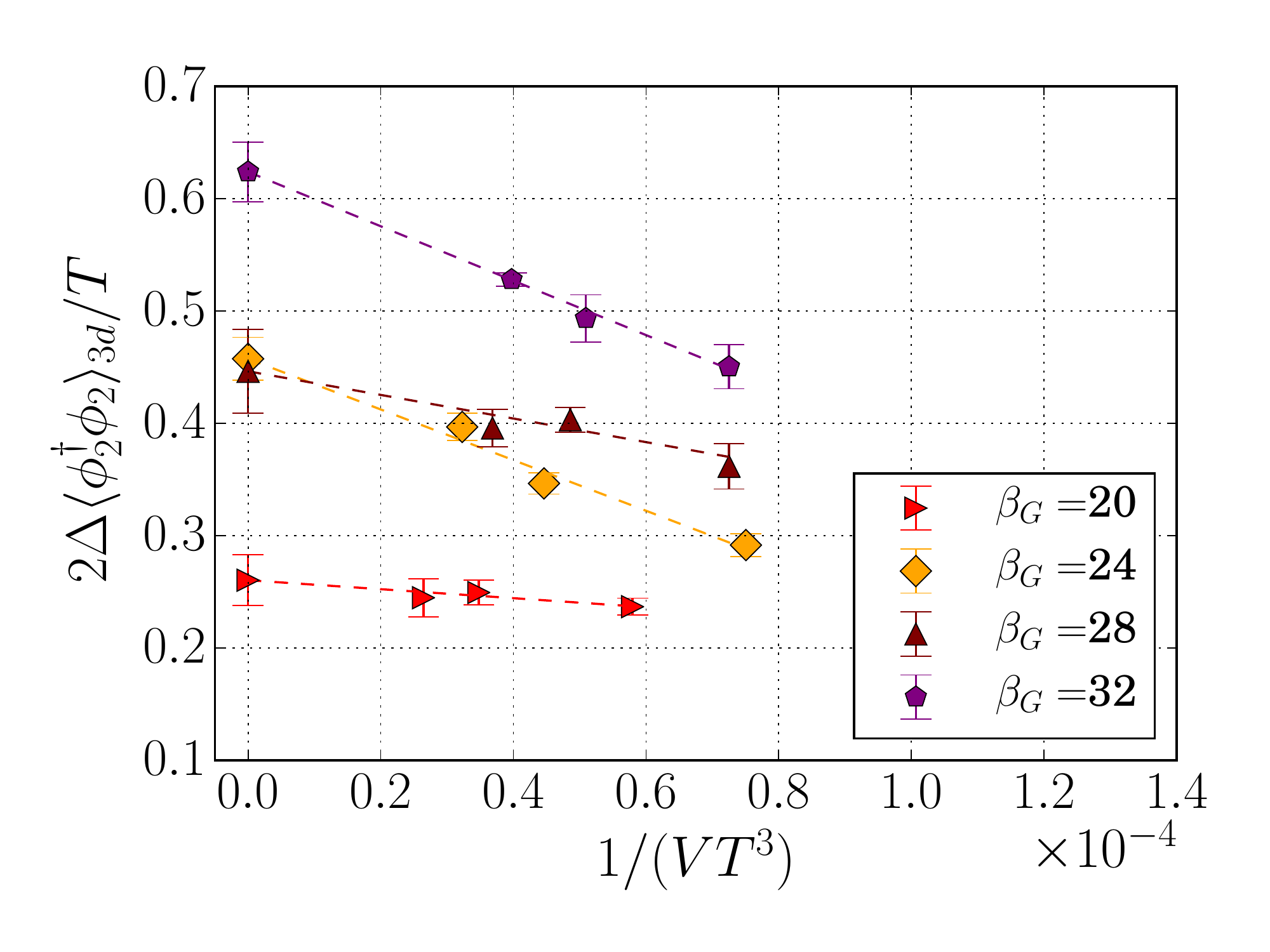}\label{}}
\caption{Volume dependence of the order parameter discontinuity $2\Delta\langle\phi_2^\dagger \phi_2\rangle_\text{3d}/T$ for different lattice spacings. }
\label{fig:orderparam_vol}
\end{figure}

From probability distributions at the critical temperature, such as those shown in Fig.~\ref{fig:BM1_hgram}, it is straightforward to measure the discontinuity in doublet condensates. In both of our BM points, however, only the $\Phi_2$ condensate can be used as an order parameter, related to the condensate in \MSbar renormalization as described in Ref.~\cite{Laine:1997dy}. Volume dependence of the dimensionless quantity $2\Delta\langle\phi_2^\dagger \phi_2\rangle_\text{3d}/T$ is shown in Fig.~\ref{fig:orderparam_vol}. Unlike in the MSSM study of Ref.~\cite{Laine:2000rm}, there is a significant dependence on the lattice volume, and we find the dependence to be approximately linear in the dimensionless combination $1/(VT^3)$.

\begin{figure}[h]
	 \subfloat[BM1]{\includegraphics[scale=0.3725]{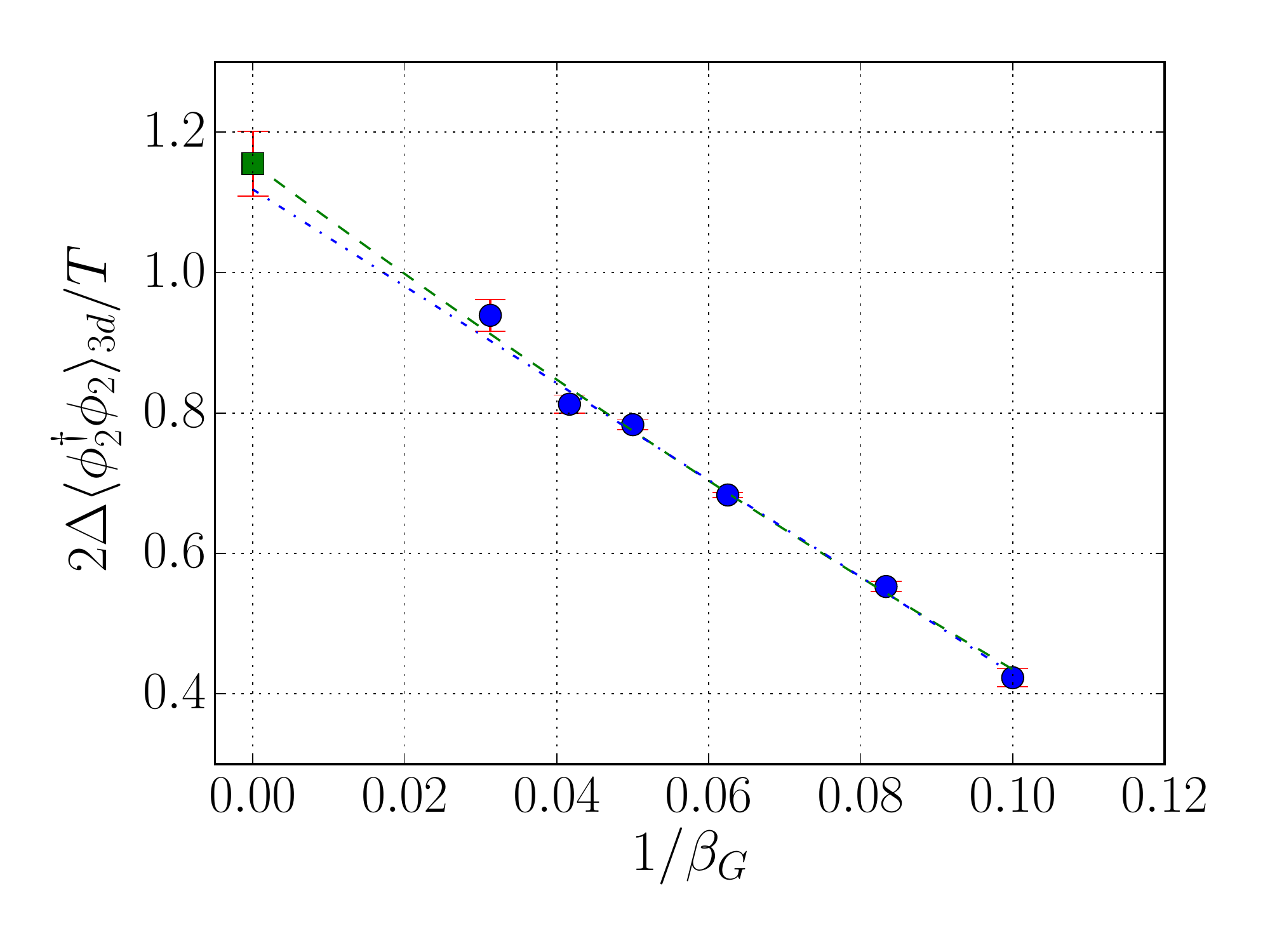}\label{}}
		\kern-0.5em
     \subfloat[BM2]{\includegraphics[scale=0.3725]{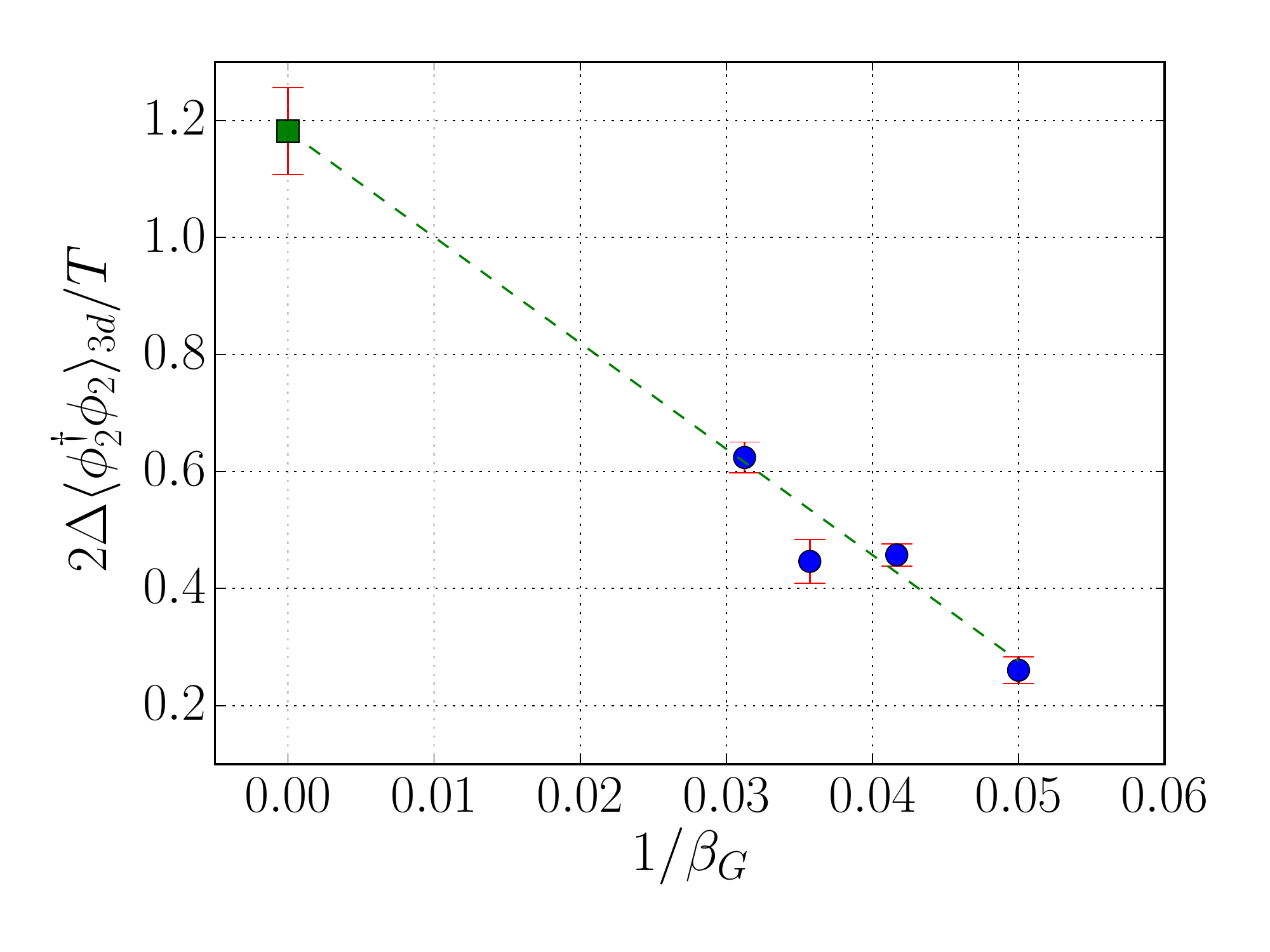}\label{}}	
     \caption{Continuum extrapolations of the order parameter discontinuity in the infinite-volume limit. Both quadratic (green) and linear (blue) fits are shown in BM1. A reliable higher-order polynomial fit in BM2 would require the use of very large $\beta_G$ -- and consequently extremely large lattices -- due to the very massive scalar degrees of freedom.}
	 \label{fig:orderparam_beta}
\end{figure}

$\beta_G$ dependence of the infinite-volume results is plotted in Fig.~\ref{fig:orderparam_beta}. In BM1, a least-squares quadratic extrapolation fits nicely to the data points, but we have also included a linear fit. The difference between the two extrapolations is roughly $3\%$ in the continuum limit, with the quadratic fit resulting in a slightly larger error. We shall use the quadratic extrapolation from now on. 

In BM2, however, extrapolation is more difficult due to the smaller range of $\beta_G$ used in the simulations. As a result, a quadratic polynomial does not fit the points well. We have therefore used a linear fit in BM2, but emphasize that missing higher-order terms may have a numerical impact. Reliably probing the effect of $1/\beta_G^2$ and higher terms would require the use of very small lattice spacings, which is computationally expensive due to the large lattices required. Given that the perturbative uncertainty from dimensional reduction and zero-temperature renormalization is possibly considerably large, of the order $20\%$ as estimated in section~\ref{sec:DR-validity}, we shall be content with a linear extrapolation here.  

Often, the quantity used for determining the strength of the EWPT is not the order parameter discontinuity $\Delta\langle\phi^\dagger \phi\rangle$, but the discontinuity in the (gauge-dependent) doublet VEV 
\be 
\langle \phi \rangle = \frac{1}{\sqrt{2}} \begin{pmatrix} 0 \\ \varphi \end{pmatrix}
\ee
-- or, in the case of multiple doublets, some combination of their VEVs -- which can conveniently be measured from the effective potential. We define a gauge-invariant counterpart to the conventional $\phi_c/T_c$ as 
\begin{align} \label{eq:order_param}
\Delta \phi/T = \Big[ \frac{2\Delta \langle\phi_2^\dagger \phi_2\rangle_\text{3d}}{T} \Big]^{\frac12},
\end{align}
which is dimensionless due to the different normalization of fields in the effective theory.

\paragraph{Latent heat:} 

By definition, the latent heat $L$ is the discontinuity in the energy density of the system, and is a concrete physical quantity characterizing the strength of the phase transition. In terms of the partition function,
\begin{align} \label{eq:latent_partition}
\frac{L}{T^4} = \frac{1}{V T^2} \Delta \left( \frac{\partial}{\partial T} \ln Z \right),
\end{align} 
which can be determined from changes in the expectation values of composite field operators. Following Ref.~\cite{Laine:2000rm}, we measure the above quantity directly on the lattice and extrapolate the results to the continuum. Unlike in the MSSM case of Ref.~\cite{Laine:2000rm} where all couplings were very small, we find the temperature dependence of quartic couplings to be significant and hence include also the interaction-term expectation values in the evaluation. Schematically, 
\begin{align}
\frac{L}{T^4} =& -\frac{1}{V T^2} a^3 \Delta\biggl\langle \Phi_1^\dagger\Phi_1 \dv{m^2_{11}}{T}  + \Phi_2^\dagger\Phi_2 \dv{m^2_{22}}{T}  + \Phi_1^\dagger\Phi_2 \dv{m^2_{12}}{T} + \text{h.c.}  \nonumber \\
& + (\Phi_1^\dagger\Phi_1)^2 \dv{\bar{\lambda}_1}{T} + \dots \biggr\rangle,
\end{align}
where the ellipsis represent the remaining interaction terms. The above quantity is easily obtained by reweighting.

\begin{figure}[h]
     \centering
     \subfloat[BM1]{\includegraphics[scale=0.3725]{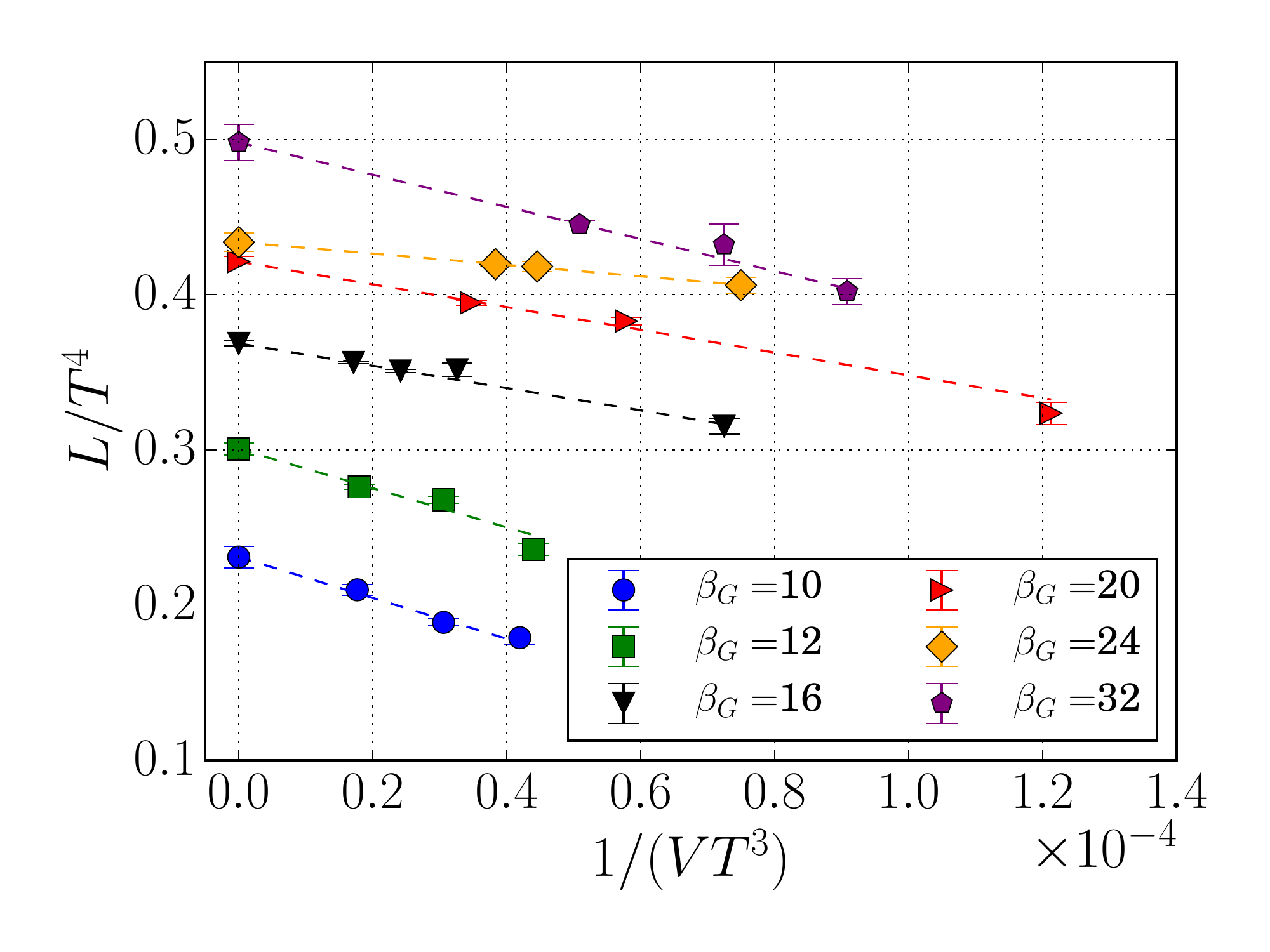}\label{}}
		\kern-0.5em
     \subfloat[BM2]{\includegraphics[scale=0.3725]{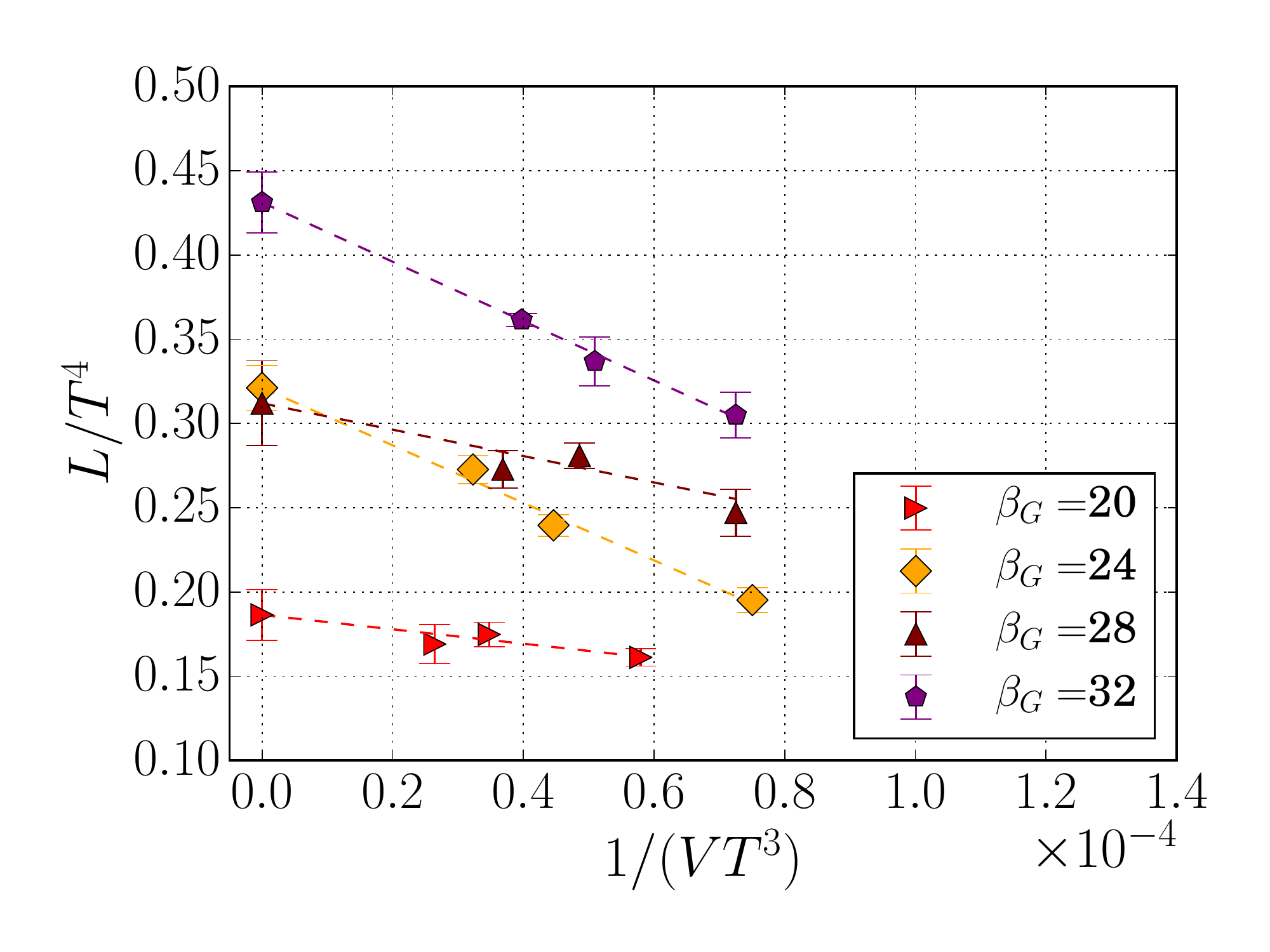}\label{}} \\
		\caption{Infinite-volume extrapolation of the latent heat.} 
		\label{fig:L_vol}
	 \subfloat[BM1]{\includegraphics[scale=0.3725]{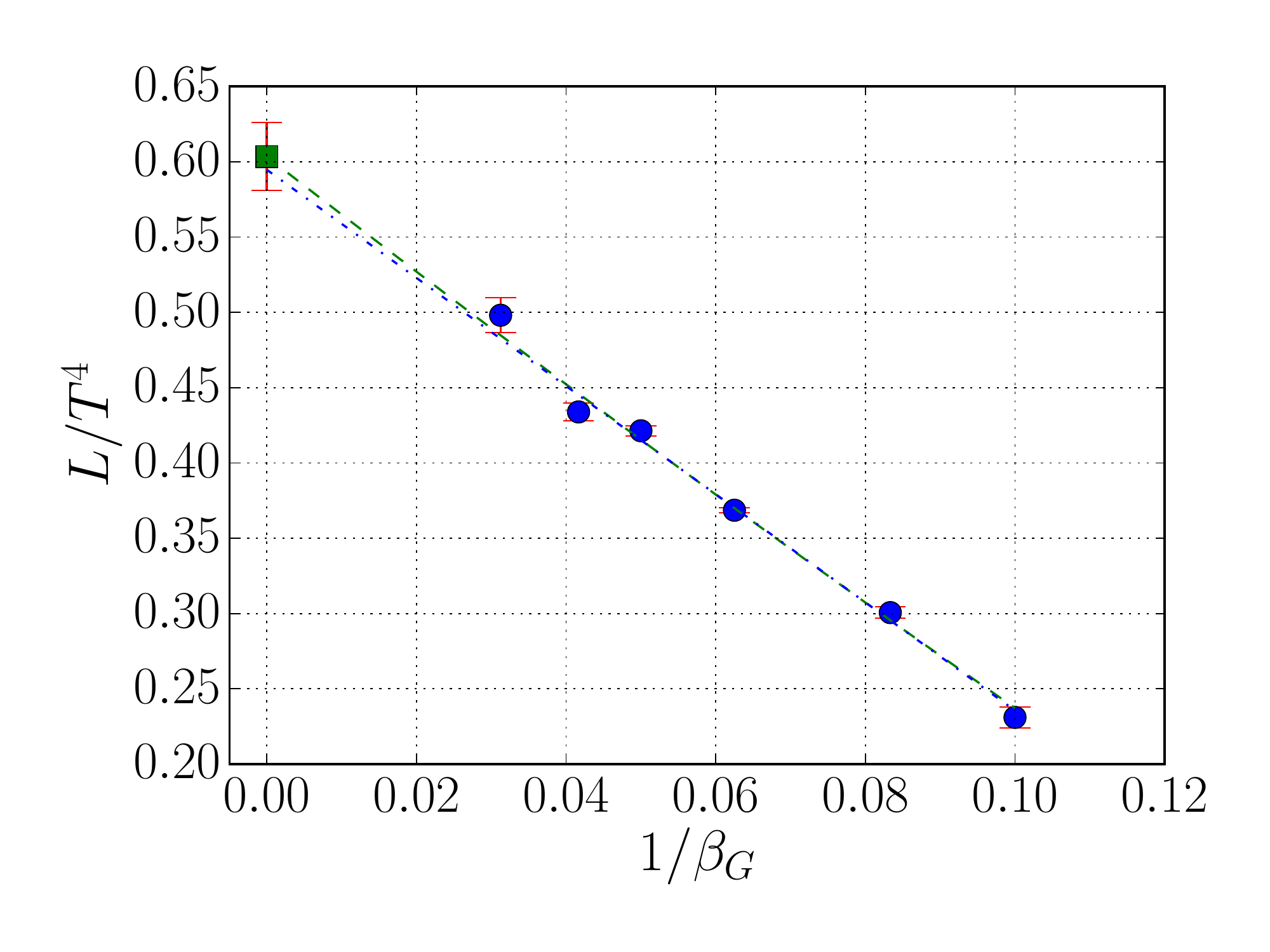}\label{}}
		\kern-0.5em
     \subfloat[BM2]{\includegraphics[scale=0.3725]{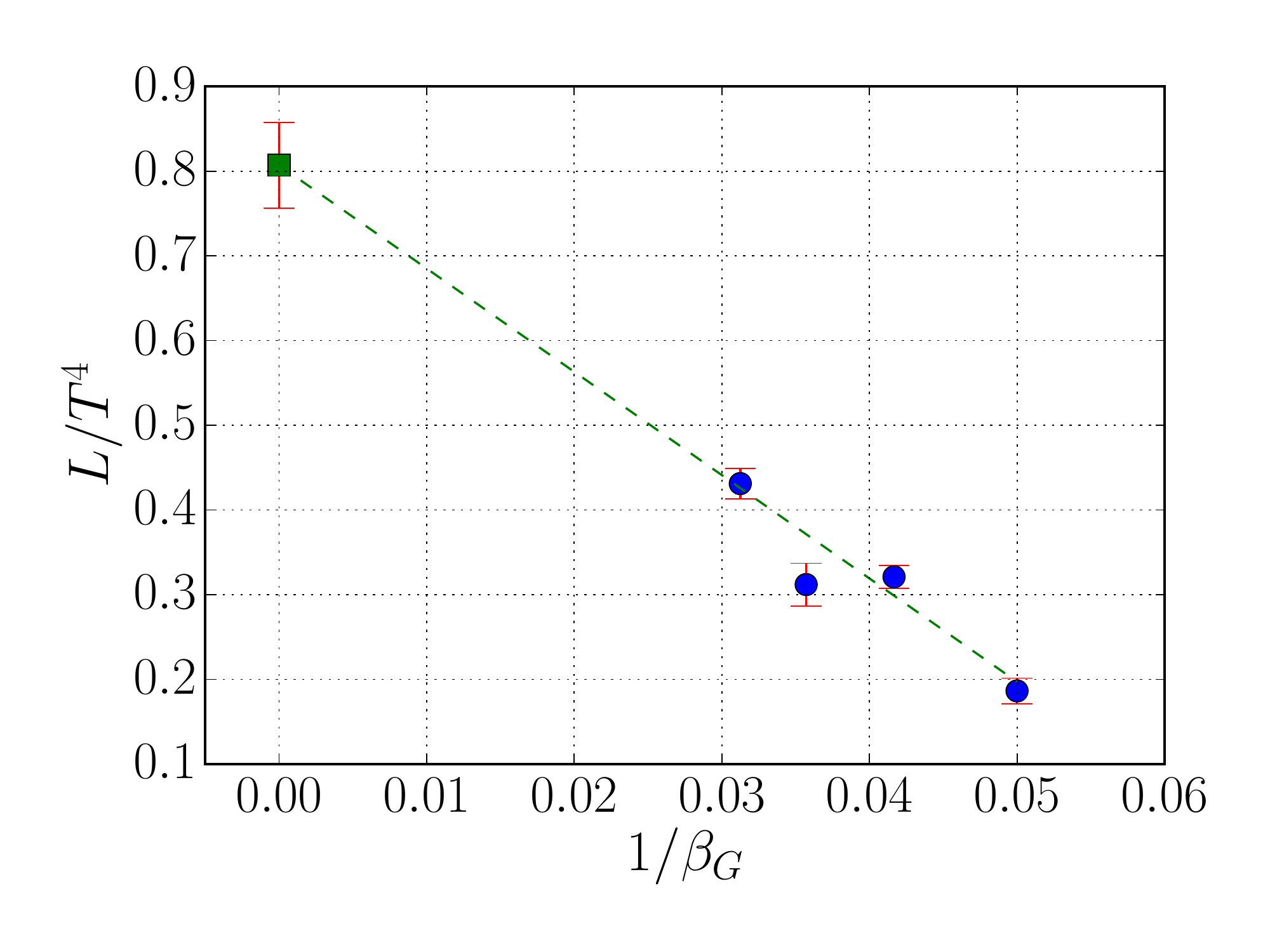}\label{}}	
     \caption{Continuum extrapolation of the latent heat. The behavior is similar to that of $\Delta v/T$, but with slightly larger statistical errors. In BM1, we again show both linear and quadratic fits.}
	 \label{fig:L_beta}
\end{figure}

Extrapolations of the latent heat to infinite volume and zero lattice spacing are shown in Figs. \ref{fig:L_vol} and \ref{fig:L_beta}. We find that the numerical behavior of the latent heat is very similar to that of the order-parameter discontinuity and have hence used the same fitting ansatzes as for $\Delta v/T$. Again in BM1, we will use the quadratic fit as the final result. As pointed out already in Ref.~\cite{Laine:2000rm}, statistical errors are somewhat larger for the latent heat. This is a natural result as there are more condensates involved in the determination of $L/T^4$.

\paragraph{Surface tension:} 

The final equilibrium quantity we measure is the tension $\sigma$ of the phase boundary separating the symmetric and broken phases. The interface tension reduces the likelihood of mixed-phase configurations, visible in the probability distributions of Fig.~\ref{fig:BM1_hgram} as a suppressed "valley" between the two phases. The suppression is proportional to $\exp(-\sigma A/T)$, with $A$ being the area of the phase boundary, and can be measured from the probability distributions using the histogram method~\cite{Iwasaki:1993qu}. Specifically, the quantity 
\begin{align}
\label{eq:hgram_method}
\frac{1}{2A} \ln\frac{P_\text{max}}{P_\text{min}},
\end{align}
where $P_\text{max}$ and $P_\text{min}$ denote the maximum and minimum probability distribution between the peaks, respectively, will tend to $\sigma/T$ in the infinite-volume limit.

In cylindrical lattices with $L_z \gg L_x = L_y$, the phase interface generally will form perpendicular to the $z$ direction, as this configuration is energetically favored over other possibilities. As in Refs.~\cite{Kajantie:1995kf,Laine:1998vn,Laine:2000rm}, we apply a finite-volume scaling ansatz in order to reduce large-volume effects related to lattice geometry. For $L_x = L_y$, an appropriate ansatz is~\cite{Iwasaki:1993qu}
\begin{align}
\label{eq:scaling_ansatz}
\frac{\sigma}{T} = \frac{1}{2(a L_x)^2} \ln\frac{P_\text{max}}{P_\text{min}} + \frac{1}{(a L_x)^2} \left[ \frac34 \ln L_z - \frac12 \ln L_x + \frac12 G + \text{const.} \right],
\end{align} 
where $G=0$ for cylindrical lattices. A periodic lattice will contain two interfaces, and it is assumed that their mutual interactions can be neglected in Eq.~(\ref{eq:scaling_ansatz}). In practice, this condition is fulfilled for long lattices where the interfaces form far enough from each other. Our lattices generally have $L_z \approx 4 L_x$ and, in this regard, are more ideal than the lattice shapes previously used in EWPT simulations. 

\begin{figure}[H]
     \centering
     \subfloat[BM1]{\includegraphics[scale=0.3725]{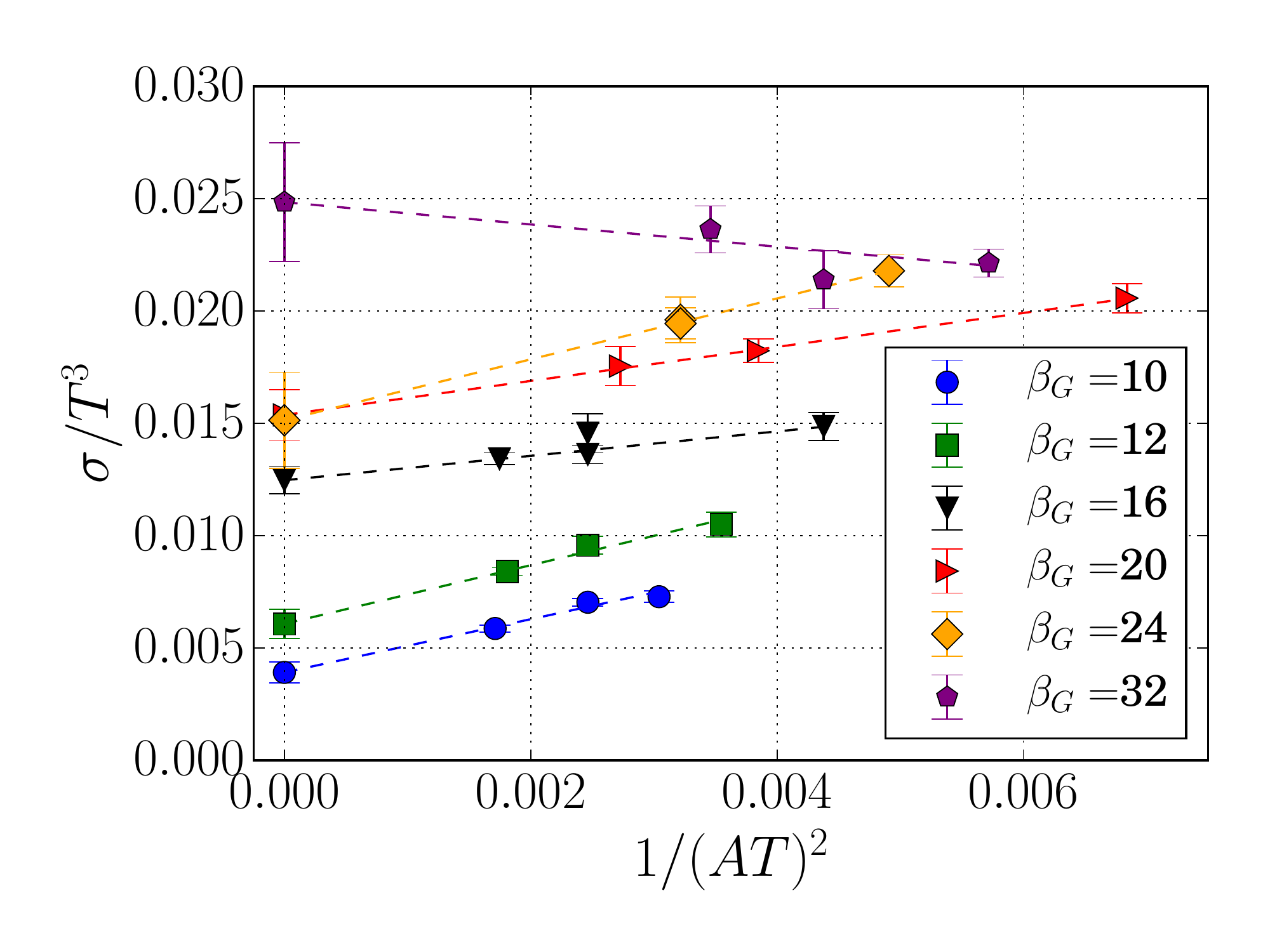}\label{}}
		\kern-0.5em
     \subfloat[BM2]{\includegraphics[scale=0.3725]{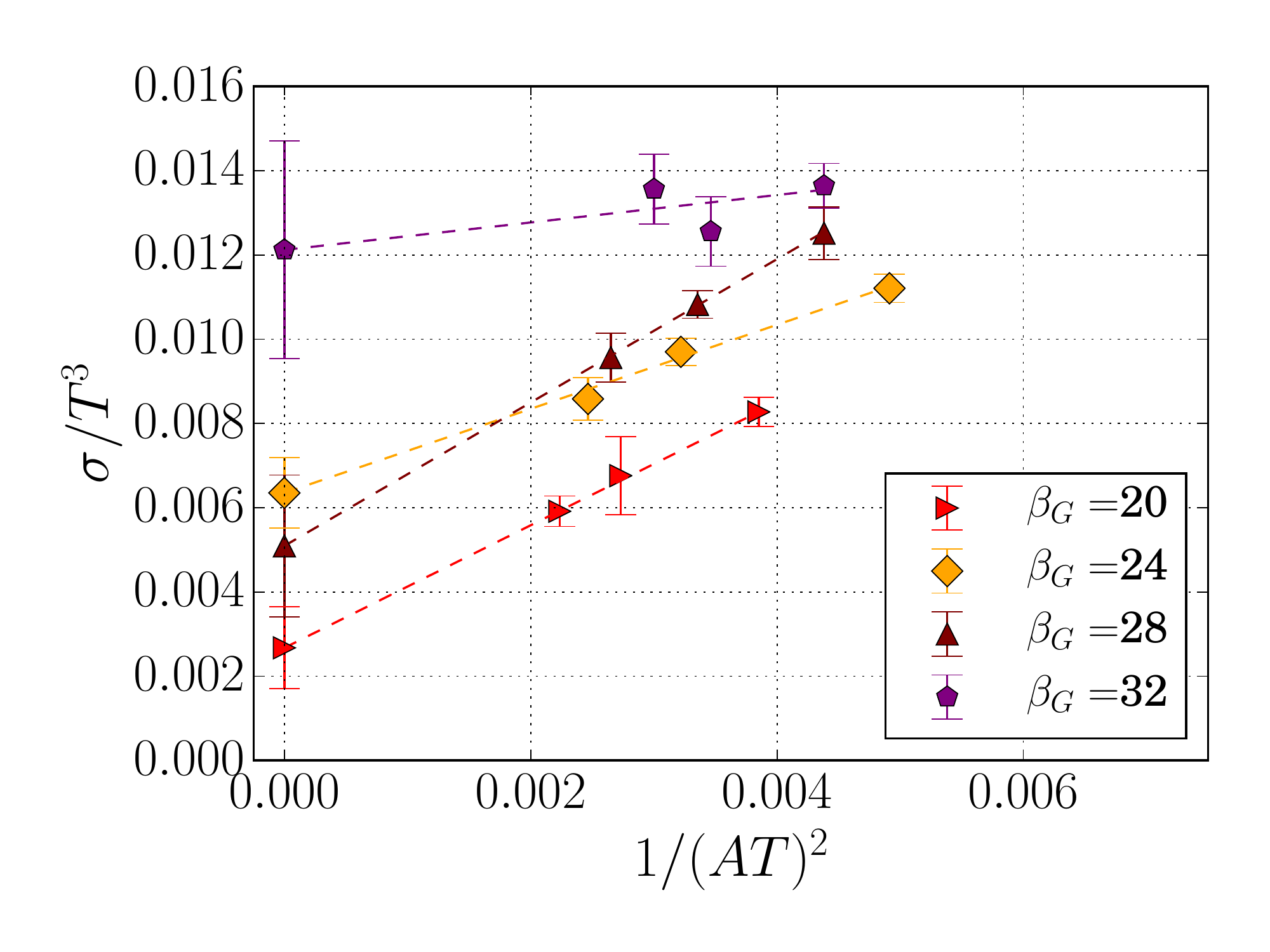}\label{}} \\
		\caption{Extrapolation of the surface tension to an infinite surface area. Statistical errors are substantial in the $\beta_G = 32$ case and could be improved with more simulations.}
		\label{fig:sigma_area}
\end{figure}

The extrapolations $1/(AT^2) \rightarrow 0$ for the dimensionless combination $\sigma/T^3$ are shown in Fig.~\ref{fig:sigma_area}. The area dependence in both BM points is linear for all of our lattice spacings, but the extrapolations for $\beta_G=32$ come with large statistical uncertainty. Improving the fits would require the use of very large lattices, being computationally expensive, and given the limited use of the surface tension in practical applications we have chosen not to pursue better accuracy here.   

\begin{figure}[H]
	 \subfloat[BM1]{\includegraphics[scale=0.37]{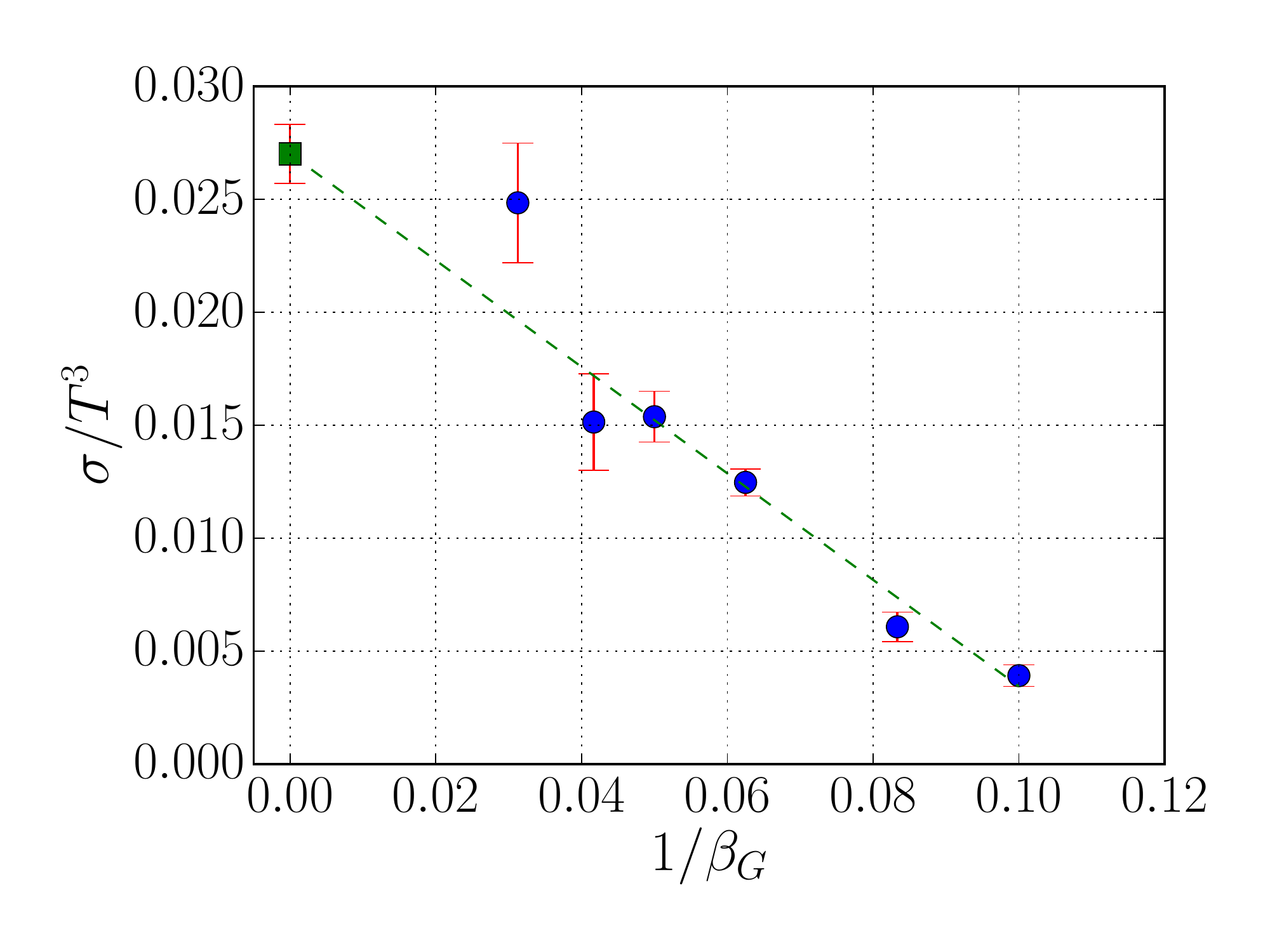}\label{}}
		\kern-0.5em
     \subfloat[BM2]{\includegraphics[scale=0.37]{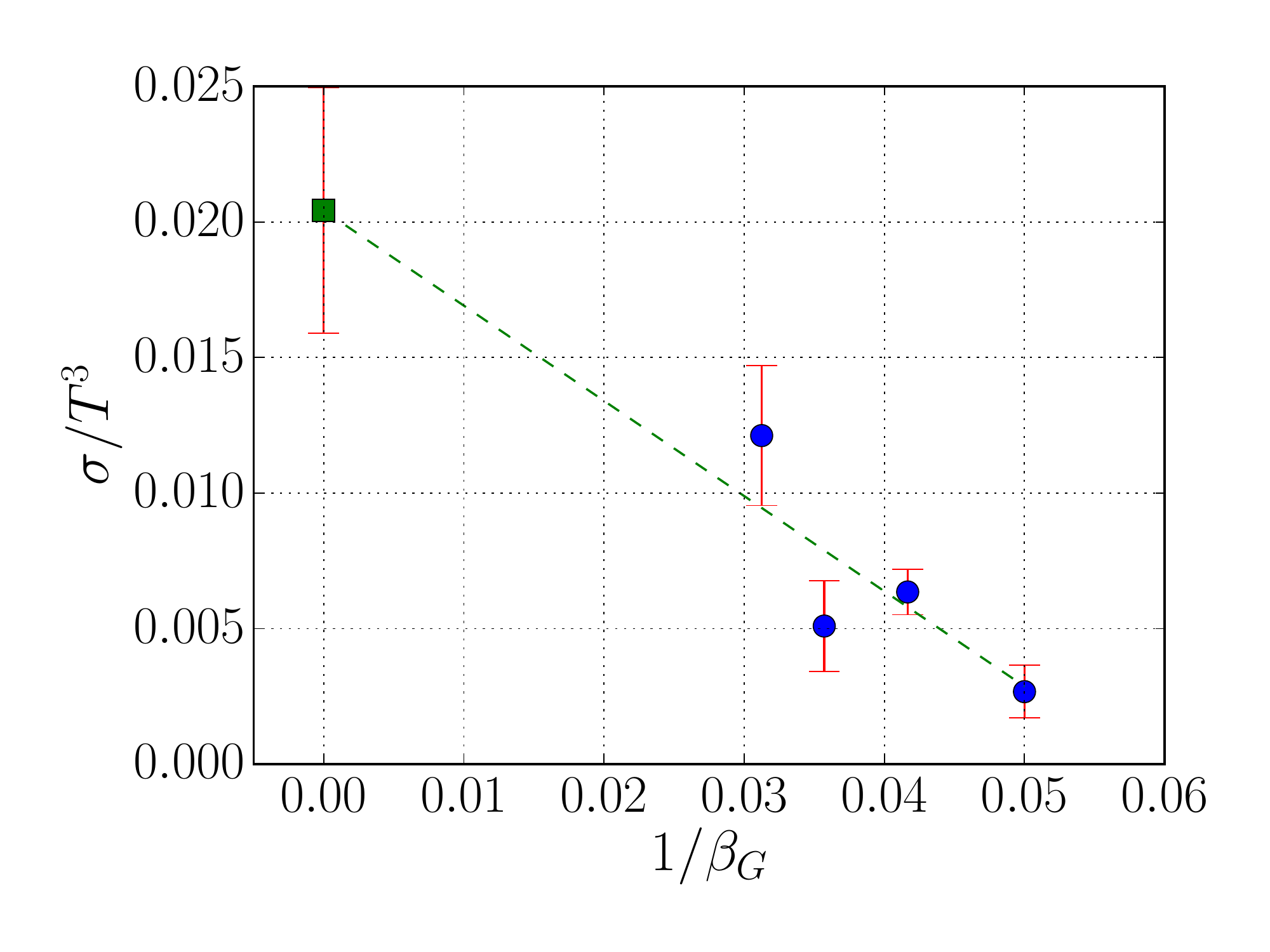}\label{}}	
     \caption{Continuum extrapolation of the interface tension.}
	 \label{fig:sigma_beta}
\end{figure}

Fig.~\ref{fig:sigma_beta} shows a continuum estimate of the surface tension. Due to the large uncertainty in the $\beta_G = 32$ infinite-volume extrapolation, we again have chosen a linear fit. Somewhat surprisingly, we find the dependence on $\beta_G$ to be roughly as important as the dependence on the interface area. This is due to the heavy degrees of freedom which become dynamical only at small lattice spacings.

\section{Comparison with perturbation theory}
\label{sec:pert_theory}

Having obtained the equilibrium characteristics of the EWPT nonperturbatively on the lattice, we now wish to evaluate the same quantities in perturbation theory and compare the results. We shall use the effective potential $\Veff(\varphi_1, \varphi_2)$ for classical background fields $\varphi_i$ and assume that the background is only in the neutral component, {\em i.e}, the doublets $\phi_i$ are shifted as 
\be
\phi_i \rightarrow \phi_i + \frac{1}{\sqrt{2}} \begin{pmatrix}  0 \\ \varphi_i \end{pmatrix} \,.
\ee

The background fields modify mass eigenvalues and couplings of the theory. The field-dependent masses of the scalars are given by the eigenvalues of the mass matrix
\begin{equation}
(\mathcal{M}^2)_{ij} = \frac{\partial^2 V}{\partial\phi_i\partial\phi_j}  \,,
\end{equation}
where $V$ is the tree-level potential (Eq.~\eqref{eq:scalar_potential}) and the indices $i,j$ refer to the components of the doublets (see Eq.~\eqref{eq:fields}). Correspondingly, gauge boson masses are obtained by diagonalizing
\begin{equation}
\mathcal{M}_g^2 = \frac{\sum_j (\varphi_j)^2}{4}\left(\begin{array}{cccc}g^2 & 0 & 0 & 0 \\0 & g^2 & 0 & 0 \\0 & 0 & g^2 & g g' \\0 & 0 & g g' & g'^2\end{array}\right) \,.
\end{equation}
Finally, the field-dependent mass of the top quark in a Type I 2HDM is  
\begin{equation}
m_t^2 = \frac{y_t^2}{2} \varphi^2_2,
\end{equation} 
and we neglect other fermions from the phase-transition analysis due to their small couplings to the scalars.

At the critical temperature, the loop-corrected effective potential will have a symmetry-breaking minimum that is degenerate with the symmetric minimum at the origin, and the strength of the phase transition can be determined from the potential barrier separating the two minima. The values of the potential in the minima are gauge invariant, in accordance with the Nielsen identity, while the values of the background fields are not. However, apparent violations of the Nielsen identity can arise if the broken minimum is not solved consistently in the loop-counting sense, resulting in residual gauge dependence formally of higher order than the calculation~\cite{Laine:1994zq,Patel:2011th}. The gauge dependence can be removed by careful resummations of Goldstone modes as in \cite{Garny:2012cg}, but we are not aware of a simple way to implement general thermal resummations consistently in this setting. Another solution to gauge dependence is to compute $\Veff$ for the composite operators $\phi^\dagger_i \phi_i$ rather than for $\varphi_i$, and the thermodynamical quantities can then be obtained from the potential in a gauge-invariant manner~\cite{Buchmuller:1994vy}.

In Ref.~\cite{Laine:2017hdk} -- where the Feynman-t'Hooft gauge was used -- the authors argued that ambiguities related to gauge dependence are overshadowed by the uncertainty related to higher-order corrections from the large scalar couplings. We shall therefore be content with the practical approach described at the beginning of this section, as is frequently done in the literature, and use Landau gauge for the perturbative calculations.

\subsection{Perturbative calculation in the effective theory}
\label{sec:3d-Veff}

We start with the effective potential constructed within the 3d EFT, Eq.~(\ref{eq:eff_theory}). This calculation is simpler than the conventional $\Veff$ in the full theory, both conceptionally and computationally, as thermal corrections have already been accounted for in the dimensional reduction procedure. As mentioned briefly in section~\ref{sec:3d-EFT}, this also includes thermal resummations beyond 1-loop order. 

The 3d $\Veff$ allows for a direct comparison with the results obtained from lattice simulations that is not affected by possible uncertainties related to dimensional reduction. In particular, the magnitude of nonperturbative effects related to the ``ultrasoft'' scale $g^2 T$, for which no resummation is possible, can be estimated by comparing the 3d $\Veff$ to the lattice results. As RG running in 3d starts only at 2-loop level, it is desirable to calculate the 3d $\Veff$ to two loops. We carry out the calculation in $d=3-2\epsilon$ spatial dimensions using the \MSbar scheme. Since the $\gr{U(1)}$ subgroup has been left out from the lattice simulations, we choose to drop its contributions to the 3d $\Veff$ as well. However, we have also performed the analysis with the full $\gr{U(1)}$ contributions included at 2-loop level and verified that their effect on the phase transition is small in comparison to systematic uncertainties in the calculation.

For a 3d EFT containing only one Higgs doublet, the 3d $\Veff$ and a list of the relevant integrals have been presented in Ref.~\cite{Farakos:1994kx}. We extend this calculation to our EFT with two doublets. Having integrated out fermions already in the dimensional reduction, the 1-loop correction to the 3d $\Veff$ is given by the bosonic zero modes as 
\begin{equation}
\begin{aligned}
\label{eq:3d_Veff_1loop}
V_\text{eff, 1-loop}^{\text{3d}} &= 2(d-1) J_3 \big(m_W(\varphi^\text{3d}_1,\varphi^\text{3d}_2)\big) + (d-1) J_3\big(m_Z(\varphi^\text{3d}_1,\varphi^\text{3d}_2)\big) \\ &+ \sum_j J_3\big(m_j(\varphi^\text{3d}_1,\varphi^\text{3d}_2)\big) \,,
\end{aligned}
\end{equation}
where the sum runs over all scalar eigenstates, and the UV-finite integral $J_3$ is given by 
\begin{equation}
J_3(m) \equiv \frac12 \int \frac{\text{d}^d p}{(2\pi)^d} \ln(p^2 + m^2) = -\frac{m^3}{12\pi}\,.
\end{equation}
We emphasize that all parameters entering the 3d $\Veff$ are those of the 3d EFT, Eq.~(\ref{eq:eff_theory}), which themselves are functions of the renormalized parameters of the full theory and the temperature, as dictated by dimensional reduction. Consequently, the masses in Eq.~(\ref{eq:3d_Veff_1loop}) are the thermally-screened masses. Note that when the $\gr{U(1)}$ sector is neglected ($\bar{g}' = 0$), we have $m_Z = m_W$. In the Landau gauge, the $\gr{SU(2)}$ ghosts are massless and do not enter the 1-loop corrections.  

\begin{figure*}
\centering
	\includegraphics{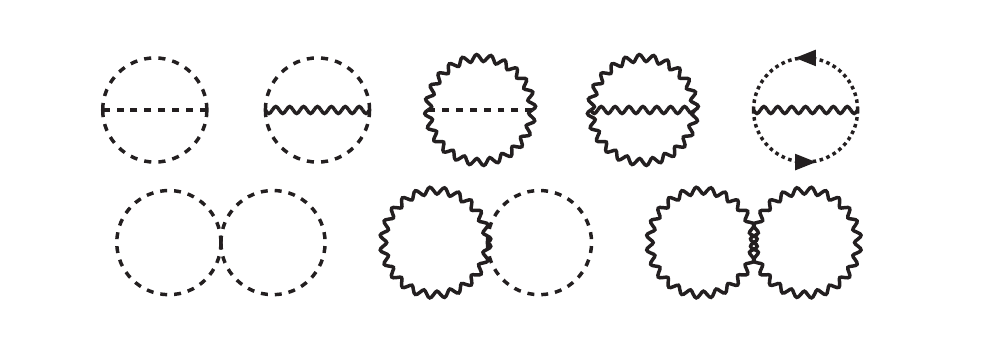}
	\caption{2-loop topologies contributing to the 3d $\Veff$, in Landau gauge. Sunset-type diagrams on the first row are divergent in the UV and exactly determine the RG evolution of the mass parameters in the 3d EFT.  }
	\label{fig:3d-twoloop}
\end{figure*}

The 2-loop correction is obtained from vacuum diagrams shown in Fig.~\ref{fig:3d-twoloop}. Although the calculation is standard, and the integrals can be found in Ref.~\cite{Farakos:1994kx}, the number of diagrams is large due to the many scalar fields present in our theory. For this reason, we choose not to give the 2-loop contribution in an explicit form, and shall simply present the numerical results obtained with the full 2-loop potential in section~\ref{sec:pert_results}.

UV divergent contributions arise at 2-loop level, which can be cancelled by introducing mass counterterms in the tree-level potential. The divergence matches the counterterms previously obtained in an independent calculation in Ref.~\cite{Gorda:2018hvi}, which serves as a cross check of our 3d $\Veff$. As a result of super-renormalizibility, the RG running of the mass parameters with respect to the \MSbar scale $\Lambda_3$ in 3d is defined exactly by the 2-loop counterterms, while the couplings remain RG invariant.

\subsection{1-loop effective potential in the full theory}
\label{sec:MSbar-Veff}

Next, we discuss the predominant approach for studying the EWPT perturbatively, namely the 1-loop resummed effective potential calculated in the full theory. It is given by
\begin{equation} \label{eq:VMS}
V_\text{eff}(\varphi_1,\varphi_2) = V(\varphi_1,\varphi_2) + V_{\rm CW}(\varphi_1,\varphi_2) + V_T(\varphi_1,\varphi_2) \,,
\end{equation}
where $V_{\rm CW}$ is the $T=0$ 1-loop Colemann-Weinberg correction to the tree-level potential $V$, and $V_T$ is the 1-loop finite temperature part. The Colemann-Weinberg part is~\cite{Coleman:1973jx}
\begin{equation} \label{eq:CW_pot}
V_{\rm CW} = \sum_j \frac{N_j}{64\pi^2} m_j^4 \left\{\ln\left[\frac{m_j(\varphi_1,\varphi_2)^2}{\Lambda^2}\right] - C_j\right\} \,,
\end{equation}
where the sum runs over the particle content of the model, $N_j$ is the internal number of degrees of freedom which is positive for bosons and negative for fermions, $m_j$ is the field dependent mass, $\Lambda$ is the renormalization scale, and $C_j=5/6$ for gauge bosons and $C_j=3/2$ for fermions and scalars. 

The unimproved thermal part is given by the integral
\begin{equation}
V_T = \frac{T^4}{2\pi^2} \sum_j N_j \int_0^\infty {\rm d}y \,y^2 \ln\left\{1\mp \exp\left[-\sqrt{y^2 + M_j^2/T^2}\right]\right\} \,.
\end{equation}
This expression can be improved by accounting for the thermal dispersion relations of the (quasi)particles. Originally Parwani~\cite{Parwani:1991gq} suggested to do this by replacing the field dependent masses of bosons by their thermal masses, $m_j \to m_j(T)$ in the whole 1-loop part of the effective potential. This approach is not self-consistent however. It leads to $T$-dependent divergences at higher orders and the choice of the thermal mass is ambiguous~\cite{Laine:2017hdk}. In a more consistent approach by Arnold and Espinosa~\cite{Arnold:1992rz}, one introduces thermal masses only in the cubic terms. This corresponds to screening only the IR-sensitive zero modes and results in the following ring-improved potential
\begin{equation} \label{eq:AE-resum}
V_{T,{\rm A-E}} = V_T + \frac{T}{12\pi} \sum_{j\in {\rm bosons}} \left[m_j^3 - m_j(T)^3\right]\,.
\end{equation}
On the other hand, this technique relies on the high-$T$ expansion in separating the contributions of the heavy modes from those of the zero-modes. Much like dimensional reduction, this resummation procedure fails when the bosonic zero modes are also heavy. For this reason, the Parwani resummation allows a smoother continuation to the nonrelativistic limit in theories which contain heavy degrees of freedom~\cite{Cline:2011mm, Laine:2017hdk}. We shall perform our numerical analysis using both of these resummation methods.

Fermions and transverse gauge boson modes do not receive thermal corrections. For the longitudinal gauge bosons, the thermal masses are obtained by diagonalizing $\mathcal{M}_g^2 + \delta \mathcal{M}_g^2$, where
\begin{equation}
\delta \mathcal{M}_g^2 = 2T^2 \left(\begin{array}{cccc}g^2 & 0 & 0 & 0 \\0 & g^2 & 0 & 0 \\0 & 0 & g^2 & 0 \\0 & 0 & 0 & g'^2\end{array}\right) \,,
\end{equation}
and the thermal scalar boson masses are obtained from $\mathcal{M}^2 + \delta \mathcal{M}^2$, where
\begin{equation}
\delta (\mathcal{M}_n^2)_{ij} = \frac{T^2}{24} \sum_k c_k N_k \frac{\partial^2 m_k^2}{\partial\phi_i \partial\phi_j} \,,
\end{equation}
with $c_k=1$ for bosons and $c_k=-1/2$ for fermions.

\subsection{Numerical results in perturbation theory}
\label{sec:pert_results}

The condition that the symmetry-breaking minimum becomes degenerate with the minimum at the origin determines the critical temperature $T_c$ and the critical field value $\phi_c \equiv \sqrt{\varphi^2_{1,c} + \varphi^2_{2,c}}$. Discontinuity in the quantity $\phi_c/T_c$ then corresponds roughly to the order parameter discontinuity obtained from lattice simulations (Eq.~(\ref{eq:order_param})), aside from the ambiguities related to gauge fixing. In the 3d EFT where the fields are scaled to mass dimension $\text{GeV}^{1/2}$, the corresponding quantity is $\phi^\text{3d}_c/\sqrt{T_c}$.

In the thermodynamic limit, the value of $\Veff$ in its minimum coincides with the grand canonical free energy density. Hence, the latent heat can be obtained from the effective potential as 
\begin{equation}
L = T_c \left( \frac{\partial V_{\text{eff}}(T,\varphi_{1,c},\varphi_{2,c})}{\partial T}\Big|_{T=T_c} - \frac{\partial V_{\text{eff}}(T,0,0)}{\partial T}\Big|_{T=T_c} \right)\,.
\end{equation}
In the 3d analysis, the factor $1/T$ multiplying the 3d action has been absorbed into the definition of $V^{\text{3d}}_{\text{eff}}$, and so the latent heat is obtained from the 3d effective potential as 
\begin{equation}
L = T^2_c \left( \frac{\partial V^\text{3d}_{\text{eff}}(T,\varphi^\text{3d}_{1,c},\varphi^\text{3d}_{2,c})}{\partial T}\Big|_{T=T_c} - \frac{\partial V^\text{3d}_{\text{eff}}(T,0,0)}{\partial T}\Big|_{T=T_c} \right)\,.
\end{equation}
For simplicity, we shall not compute the surface tension perturbatively.

The effective potential bears and explicit dependence on the RG scale $\Lambda$. While the full effective action is $\Lambda$-independent, in perturbative expansions there is always an uncertainty related to the scale variation, higher in order than the one under consideration. Due to the large scalar couplings present in our analysis, this ambiguity in the choice of $\Lambda$ is a significant source of uncertainty. We demonstrate this by varying the RG scale along a range of mass scales with dominant contributions to the effective potential. Clearly the most dangerous logarithms are those proportional to the scalar couplings. In the full theory, contributions of the form $\lambda_i^2\ln(m^2/\Lambda^2)$, where $m$ denotes a scalar mass, arise from the $T=0$ loop corrections, Eq.~(\ref{eq:CW_pot}). However, thermal fluctuations generate additional logarithms of the type $\ln(\Lambda/(\pi T))$, making it difficult to ensure that all logarithmic corrections stay small simultaneously. We choose to vary $\Lambda$ between $0.5\pi T$ and $1.5\pi T$.

The situation is different in the 3d EFT, where the 3d $\Veff$ contains only logarithms of the type $\ln(m_3^2 / \Lambda_{3}^2)$, which furthermore arise only at 2-loop level at the earliest. Having integrated out the mass scale $\pi T$, we vary the RG scale of the 3d EFT, $\Lambda_3$, between $0.5 T$ and $2T$. However, scale variations in the full theory affect the parameters of the EFT via the matching relations obtained from dimensional reduction. Although this uncertainty is less severe than the corresponding ambiguity in the full $\Veff$ as only thermal logarithms appear in dimensional reduction. For the sake of having a one-to-one comparison of perturbative and nonperturbative analyses in the 3d EFT, we fix the RG scale of dimensional reduction as in Eq.~(\ref{eq:DR-scale}), but discuss variations of this scale further in section~\ref{sec:DR-validity}.

\begin{figure*}
  \begin{center}
    \subfloat[BM1]{
	  \includegraphics[width=0.33\textwidth]{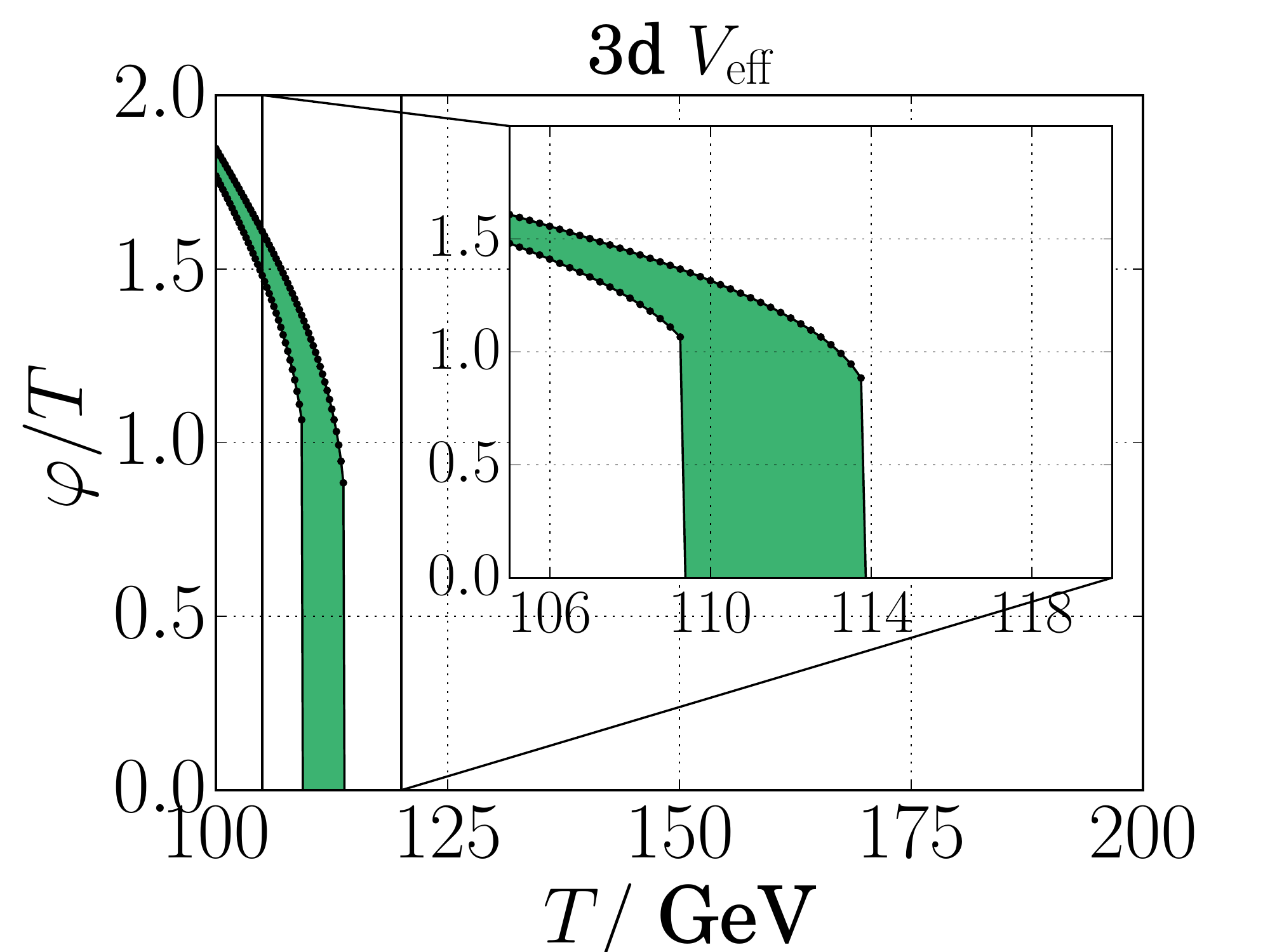}
      \includegraphics[width=0.33\textwidth]{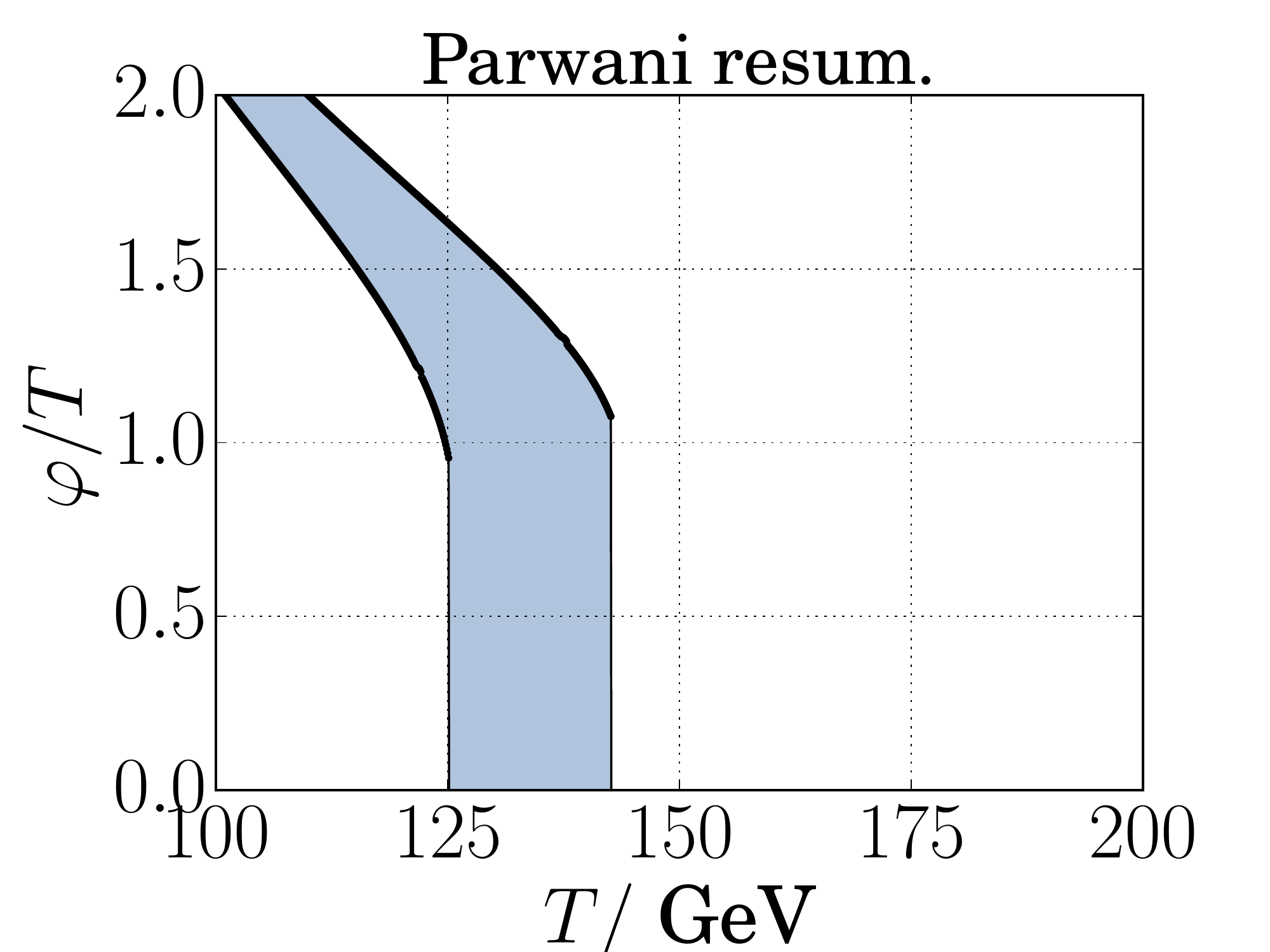} 
	  \includegraphics[width=0.33\textwidth]{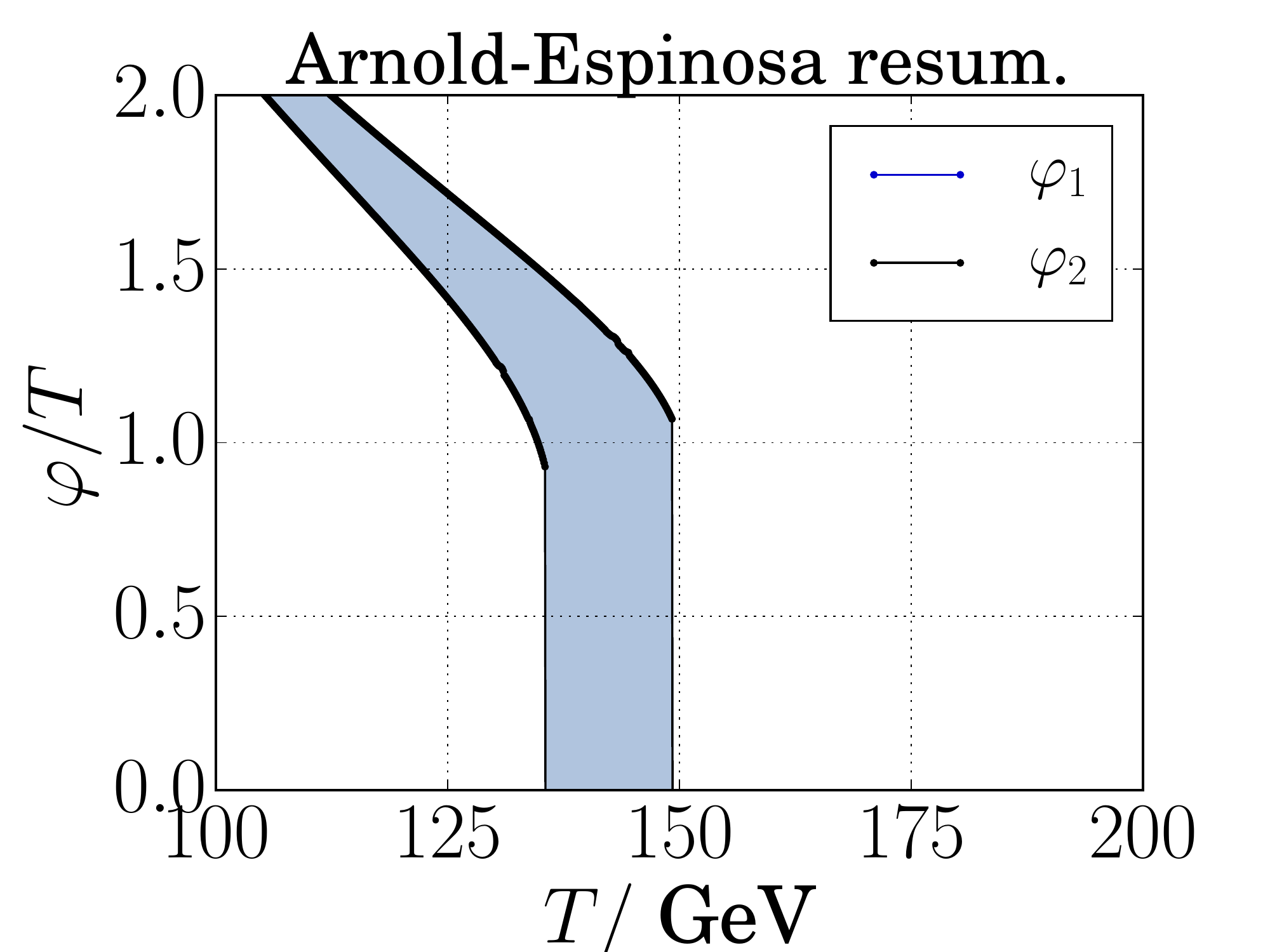}
	} \\
\vspace{-0.5cm}
    \subfloat[BM2]{
      \includegraphics[width=0.33\textwidth]{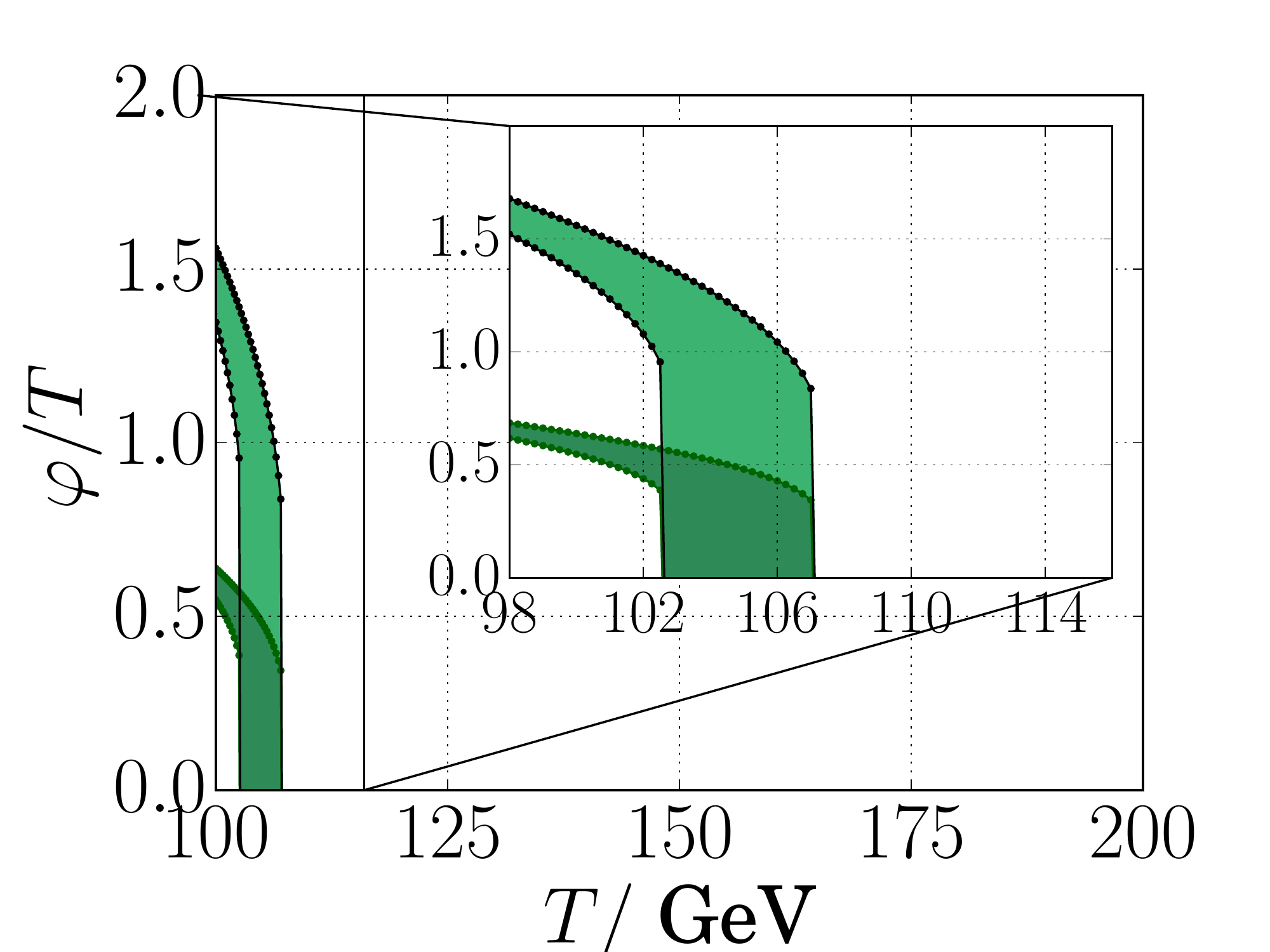}
      \includegraphics[width=0.33\textwidth]{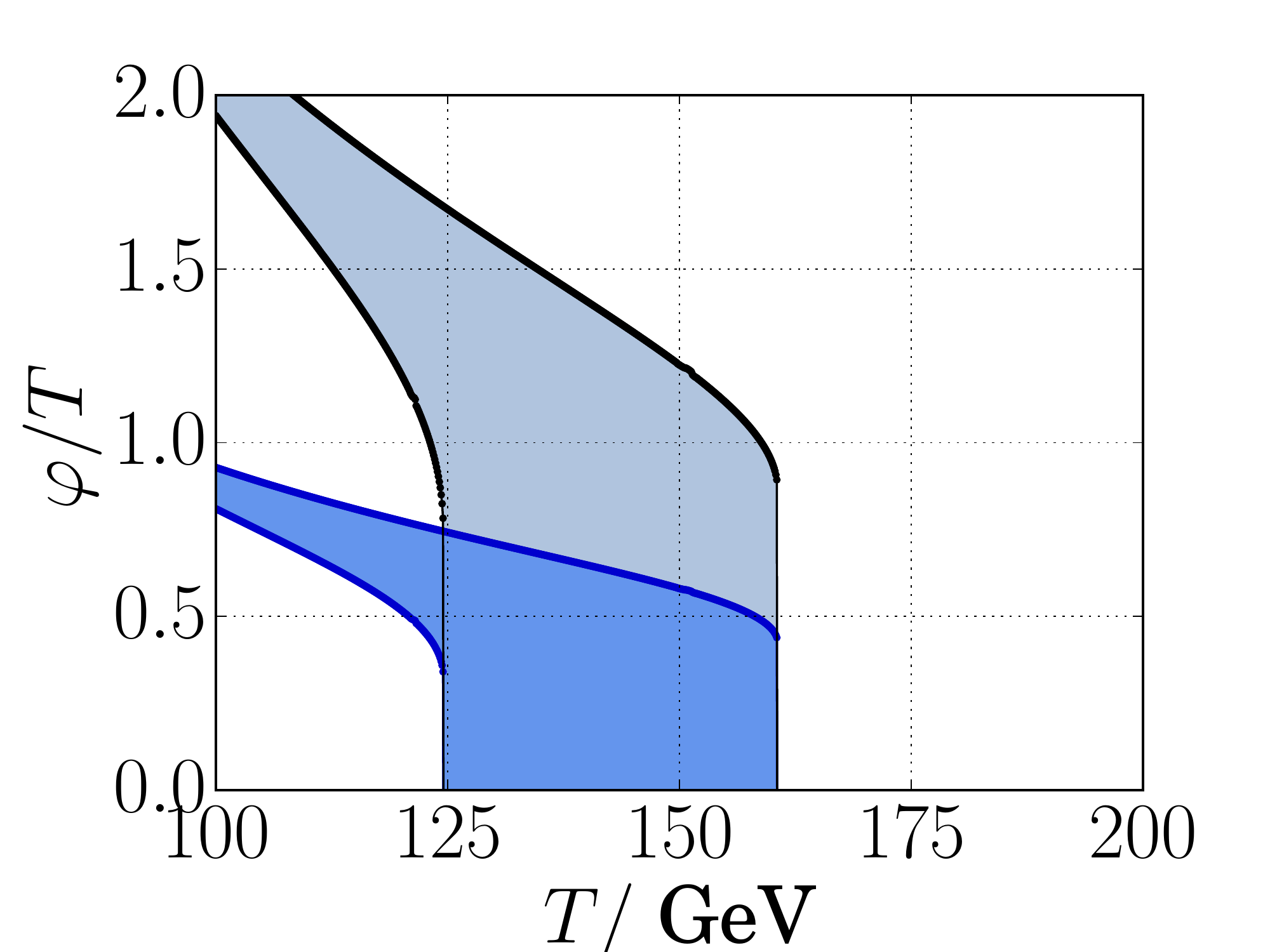} 
	  \includegraphics[width=0.33\textwidth]{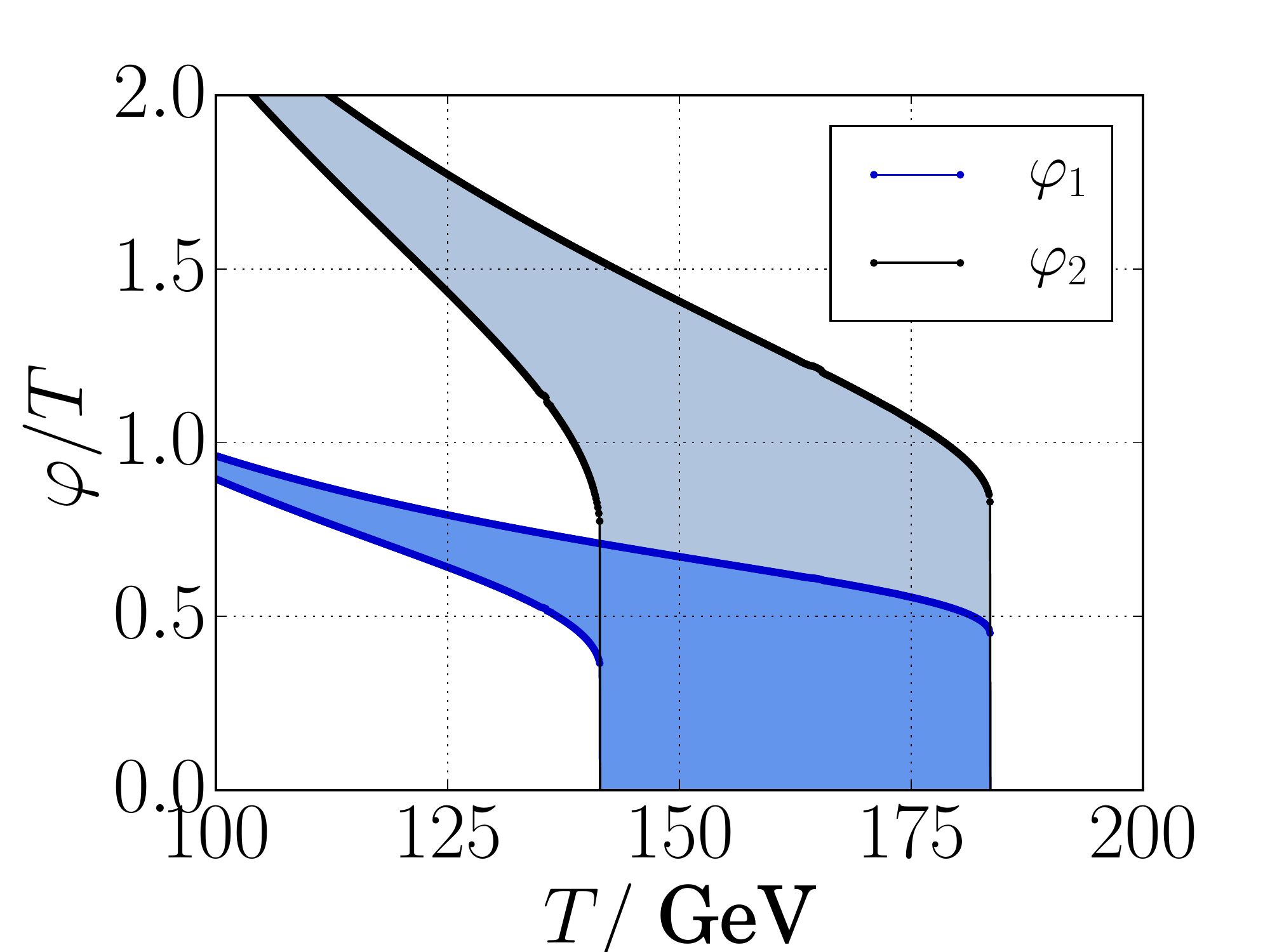}
	} \\
  \end{center}
  \caption{Location of the global minimum of the perturbative effective potential as a function of the temperature. At high temperature, the minimum is at the origin and the electroweak symmetry is restored. From left to right: 2-loop potential in the 3d EFT, 1-loop $\Veff$ with Parwani resummation, 1-loop $\Veff$ with Arnold-Espinosa resummation. Values for two different choices of the RG scale are shown, and the coloured band in between illustrates the perturbative uncertainty as described in the text. In BM1, the inert doublet $\phi_1$ does not develop a VEV due to unbroken $Z_2$ symmetry.}
\label{fig:PT_VEVs}
\end{figure*}

Turning to numerical analysis, we present our findings in Fig.~\ref{fig:PT_VEVs} and Table~\ref{table:PT_comparison}. In Fig.~\ref{fig:PT_VEVs}, we have plotted the global minimum of the potential as a function of the temperature with different resummation implementations, as well as in the 3d EFT. For each scenario, the RG scale has been varied as described above and the results for the smallest and largest scale are plotted in Fig.~\ref{fig:PT_VEVs}. The colored bands depict the uncertainty related to the scale sensitivity. In all plots, the parameters given in Table~\ref{table:input-params} have been run to the final scale using 1-loop $\beta$ functions in the full theory. In the 3d analysis, the corresponding parameters in the 3d EFT are first obtained from dimensional reduction, and then run to the final 3d scale using exact RG evolution.

For both Parwani and Arnold-Espinosa resummations in the full theory, there is significant uncertainty in determination of critical temperature, with Parwani resummation giving a considerably smaller $T_c$. On the contrary, the magnitude of the jump $\phi_c/T_c$ is not very sensitive to scale variations. In BM2, scale variations lead to much larger uncertainty than in BM1 as  the couplings and masses are larger. In 3d perturbation theory, scale uncertainties are significantly less alarming, which is not surprising as the 3d EFT is super-renormalizable and the 3d $\Veff$ is evaluated to two loops.

\begin{table*}[h!]
\begin{center}
  \begin{tabular}{l|l|l|l|l|l}
   & Method & $T_c$/GeV & $L/T^4_c$ & $\phi_c/T_c$ & $L/\text{GeV}^4$  \\
\hline
\multirow{4}{*}{BM1} 
    & 1-loop Parwani resum.  & $ 134.0 \pm 8.75 $ & $0.396 \pm 0.002 $  & $ 1.01 \pm 0.06 $ & $1.27 \times 10^8$  \\
    & 1-loop A-E resum.  & $ 142.4 \pm 6.88 $ & $0.33 \pm 0.02 $  & $ 1.00 \pm 0.07 $ & $1.37 \times 10^8$   \\
	& 2-loop $\Veff$ in 3d  & $111.6 \pm 2.30$ & $0.57 \pm 0.10$ &  $0.98 \pm 0.09$ & $0.89 \times 10^8$  \\
	& 3d lattice & $116.40 \pm 0.005$ & $0.60 \pm 0.02$ & $1.08 \pm 0.02$ & $1.11 \times 10^8$  \\
\hline
\multirow{4}{*}{BM2} 
    & 1-loop Parwani resum.  & $ 142.6 \pm 18.0 $ & $0.29 \pm 0.04 $  & $ 0.91 \pm 0.06 $ & $1.19 \times 10^8$  \\
    & 1-loop A-E resum.  & $ 162.5 \pm 21.0 $ & $0.20 \pm 0.03 $  & $ 0.88 \pm 0.05 $ & $1.36 \times 10^8$   \\
	& 2-loop $\Veff$ in 3d & $104.9 \pm 2.30$ & $0.61 \pm 0.10$ &  $0.97 \pm 0.06$ & $0.74 \times 10^8$  \\
	& 3d lattice & $112.5 \pm 0.01$ & $0.81 \pm 0.05$ & $1.09 \pm 0.03$ & $1.29 \times 10^8$  \\
  \end{tabular}
  \caption{Comparison of thermodynamic quantities as obtained either with the EFT approach, or with the resummed 1-loop potential in the full theory. Error bars indicate sensitivity to the RG-scale as described in the text. For the lattice results we show the statistical uncertainty.  Numbers in the last column correspond to central values without scale uncertainties. }
  \label{table:PT_comparison}
\end{center}
\end{table*}

Thermodynamic quantities obtained from the perturbative calculations are collected in Table~\ref{table:PT_comparison}, along with the nonperturbative lattice results from section~\ref{sec:lattice_results}. Comparing first the perturbative and nonperturbative results within the 3d EFT, we see that in both BM points, the 3d $\Veff$ describes the EWPT quite well. The strength of the transition is  slightly underestimated by the perturbative analysis in BM2 and there is a $\sim 5\%$ discrepancy in the critical temperature. Qualitatively the behavior is similar to that of the MSSM case~\cite{Laine:2000rm,Laine:2012jy}.

We emphasize that apart from perturbative corrections beyond two loops, the main difference between the 3d perturbative and nonperturbative approaches is the handling of the IR sensitive fields in the symmetric phase. As such, we conclude that these nonperturbative IR effects are, in fact, already suppressed for the marginally strong phase transitions considered here. This is reassuring, and suggests that the EWPT can be studied reliably with the relatively simple 2-loop 3d $\Veff$, at least in the cases considered here. 

Moving on, we find that the situation is not as good for the 1-loop potential in the full theory. In particular, this approach overestimates $T_c$ by a large margin compared to the analysis in the 3d EFT. This is because the critical temperature is particularly sensitive to thermal mass corrections and resummations, which are incorporated at 2-loop level in the 3d EFT. With additionally the scale uncertainty in $T_c$ being about 10 to 25 percent, we conclude that at least in our studied BM points, the full 1-loop $\Veff$ with thermal resummations is not a reliable tool for determining the temperature scale of the EWPT. On the other hand, $L$ and $\phi_c/T$  qualitatively match the lattice results, but the dimensionless combination $L/T_c^4$, important for gravitational-wave predictions, is underestimated due to the overly large $T_c$.

BM1 has previously been studied at the full 2-loop level without dimensional reduction in Ref.~\cite{Laine:2017hdk}, where a Parwani-type resummation was applied. According to their Table 4, the 2-loop corrections to the potential make the transition slightly weaker and therefore shift the result away from what we find in our lattice analysis. As the authors point out, their resummation scheme fails to take into account $\mathcal{O}(\lambda^2_3 \mu^2_{i})$ and $\mathcal{O}(\lambda^2_3 T^2)$ corrections to the thermal masses, which in turn are included in our 3d EFT and could explain the discrepancy. This ambiguity in resummation at 2-loop level is another reason why the 3d EFT approach is preferable over calculations in the full theory.

\subsection{Validity of the effective theory}
\label{sec:DR-validity}

Let us reiterate that while our nonperturbative analysis in the 3d EFT is exact within statistical errors of Monte Carlo methods, the derivation of the EFT by dimensional reduction is still performed perturbatively. Therefore, estimating the perturbative accuracy of the dimensional reduction is a crucial part of our comparison with fully perturbative methods. In this work, we have applied the dimensional reduction of Ref.~\cite{Gorda:2018hvi}, which is performed at 1-loop level for the couplings, and at 2-loop level for the masses. In the SM, where all couplings are small compared to the scalar couplings in our BM points, this next-to-leading order (NLO) calculation (cf. dimensional reduction at leading order meaning tree-level matching for couplings and one loop for masses) is highly accurate and reproduces the equilibrium thermodynamics of the EWPT with errors of only $1\%$ or less~\cite{Kajantie:1995dw}.  

At NLO, the couplings do not obtain significant loop corrections, while the mass parameters in the EFT -- the Debye screened masses --  have the schematic form
\begin{align} \label{eq:3d-mass}
\bar{\mu}^2 = \mu_0^2(\Lambda) + \Pi_{\text{1-loop}}(\Lambda) + \Pi_{\text{2-loop}}(\Lambda),
\end{align}
where $\mu_0^2$ is the corresponding mass parameter in the full theory and the loop corrections $\Pi_{\text{1-loop}}$ and $\Pi_{\text{2-loop}}$ are of the order $\mathcal{O}(\lambda T^2)$ and $\mathcal{O}(\lambda^2 T^2)$, respectively, with additional mass-dependent corrections coming from the high-$T$ expansion. 
When running of the parameters is taken into account, dependence on the RG scale $\Lambda$ can be shown to cancel exactly up to corrections formally of 3-loop order~\cite{Kajantie:1995dw,Gorda:2018hvi}. The phase transition occurs when at least one of the doublets in the EFT becomes very light, for which a cancellation between the tree-level mass $\mu_0^2$ and the thermal loop corrections is necessary. A qualitative description may already be obtained at leading order with just the 1-loop thermal mass, but for quantitative results the 2-loop part in Eq.~(\ref{eq:3d-mass}) can be significant, especially when some couplings are large. In BM1, the relative importance of the 2-loop corrections for mass parameters $\bar{\mu}^2_{11}$, $\bar{\mu}^2_{22}$ and $\bar{\mu}^2_{12}$ are 13, 17 and 2 percent, respectively, when the temperature is fixed to the critical value obtained from the simulations. In BM2, the respective numbers are 11, 19 and 2 percent.

Although the above numbers do not immediately signal bad convergence, it is possible for corrections at higher loop orders to be substantial. We can estimate the importance of higher-order effects of $\mathcal{O}(\lambda^3)$ and higher by varying the RG scale $\Lambda$ in Eq.~(\ref{eq:3d-mass}) and the other matching relations (see Ref.~\cite{Gorda:2018hvi}). This leads to increased uncertainty in our results within the 3d EFT, and while it is difficult to quantitatively estimate this effect on the lattice due to the computational effort required, we may use the 3d $\Veff$ to address the issue. Hence, we have repeated the perturbative analysis of the previous section for the 3d $\Veff$ with different scales used in the dimensional reduction. We chose the same range for $\Lambda$ which was used for the $\Veff$ in the full theory, $0.5\pi T$ to $1.5\pi T$, and the results are shown in Table~\ref{table:Veff-3d-vary-DR-scale}.

\begin{table*}[h!]
\begin{center}
  \begin{tabular}{l|l|l|l|l}
    & $T_c$/GeV & $L/T^4_c$ &  $\phi_c/T_c$ & $L/\text{GeV}^4$  \\
\hline
BM1: 3d $\Veff$ & $112.25 \pm 2.86$ & $0.55 \pm 0.11$ &  $0.95 \pm 0.10$ & $0.87 \times 10^8$  \\
\hline
BM2: 3d $\Veff$ & $106.63 \pm 4.03$ & $0.71 \pm 0.27$ &  $1.04 \pm 0.19$ & $0.91 \times 10^8$  \\
\end{tabular}
\caption{Results in 3d perturbation theory with the renormalization scale of dimensional reduction varied from $0.5 \pi T$ to $1.5 \pi T$, and the 3d scale varied from $0.5 T$ to $2 T$ as in Table~\ref{table:PT_comparison}. For corresponding scale uncertainties, we show the most pessimistic values in the error bars. }
  \label{table:Veff-3d-vary-DR-scale}
\end{center}
\end{table*}

In BM1, the variation of $\Lambda$ amounts to a slight increase of the uncertainty compared to the earlier case in Table~\ref{table:PT_comparison}, where $\Lambda$ was fixed as in Eq.~(\ref{eq:DR-scale}). In BM2, there is a clear shift in central values and the uncertainties are substantially larger than in the case of fixed $\Lambda$. We therefore estimate that higher order effects in the dimensional reduction procedure could lead to inaccuracies of a few percents in the thermodynamic quantities for the BM1 case, while for BM2, the uncertainty may be as large as few tens of percents, which can compromise quantitative predictions.

From the simple consideration above, it would clearly be preferable to include next-to-next-to leading order (NNLO) contributions to the dimensional reduction. For the parameters of the effective theory in Eq.~(\ref{eq:eff_theory}), this means 2-loop contributions in the couplings and 3-loop contributions in the masses, and also higher-order terms originating from high-$T$ expansions. While this calculation is highly non-trivial, important simplifications can be made by focusing only on the most dominant contributions from the large scalar couplings at order $\mathcal{O}(\lambda^3)$. Recent developments in calculating higher-order corrections to the dimensionally-reduced EFT of hot QCD (see~\cite{Ghisoiu:2015uza,Laine:2018lgj} and the references therein) motivate applying similar techniques in the context of the EWPT. However, we will not consider such improvements in this study.

Finally, at NNLO it is no longer justified to neglect higher-dimensional operators from the 3d EFT. In our case, we expect the most dominant operators to be scalar operators of the type $(\phi^\dagger \phi)^3_\text{3d}$, which in principle are straightforward to include in the EFT by matching scalar 6-point functions. Unfortunately, these higher-dimension operators ruin the super-renormalizibility of the 3d EFT, turning it into ``only'' a renormalizable theory. Consequently, lattice analyses are complicated as the relations to continuum parameters are no longer exact. We may nevertheless estimate the effect of these operators by including them in the perturbative $\Veff$. If the thermodynamic quantities obtained from this potential differ considerably from those in Table~\ref{table:PT_comparison}, we can then conclude that the higher-order operators' contributions are too significant to be ignored in the lattice simulations.

A systematic determination of the aforementioned NNLO contributions, as well as a numerical study of their effects in BM1 and BM2, will be published in a separate work. According to preliminary results of this work, one can estimate that the inclusion of higher-order operators weakens the transition by a few percentages in BM1 and about 20 percents in BM2, while the critical temperature is not affected significantly. This estimate has been obtained by performing a 1-loop dimensional reduction for all scalar operators of dimension six (ignoring operators containing derivatives) and including their effect in the 3d effective potential at a full 2-loop level.

Combined with the scale variation estimate above, we approximate that the accuracy of the dimensional reduction for thermodynamic quantities is within a few percentages in BM1. In BM2 with somewhat larger couplings, the accuracy is significantly worse, of the order $20\%$, and suggests that 3-loop corrections should be included if one seeks quantitative results. For even larger couplings -- such as those used in Refs.~\cite{Dorsch:2016nrg,Caprini:2015zlo} to produce very strong transitions in the heavy $M_A = M_{H^\pm}$ regime -- we believe that perturbation theory may fail to give even a qualitative picture of the EWPT. New techniques relying neither on the perturbative effective potential nor dimensional reduction are then required for reliable results. 

We emphasize that the shortcomings related to NNLO corrections in the dimensional reduction are also present in perturbative analyses in the full theory, such as the 1-loop $\Veff$, Eq.~(\ref{eq:VMS}), in the form of missing higher-order corrections. However, due to the 2-loop mass corrections from dimensional reduction and the efficient handling of IR resummations, our EFT approach is still superior to the frequently-used 1-loop $\Veff$ approach.

\section{Conclusions}
\label{sec:conclusion}

In this paper, we have performed a state-of-art study of the equilibrium properties of the electroweak phase transition in the Two Higgs Doublet model. The main analysis is based on a dimensionally-reduced effective theory~\cite{Kajantie:1995dw,Gorda:2018hvi}, obtained by integrating out the heavy thermal field modes while simultaneously incorporating thermal resummations beyond the leading order. Using the effective theory, we perform nonperturbative lattice simulations in two benchmark points motivated by model phenomenology. As a comparison, we also apply conventional perturbative methods to compute the critical temperature and the strength of the phase transition at 1-loop level.

In simple beyond the Standard Model settings where only one phase transition occurs near the electroweak scale, it is necessary to have sufficiently strong interactions in the scalar sector in order to produce a strong transition. We demonstrate that this requirement of large couplings substantially reduces the predictive power of traditional perturbative approaches, due to the significant corrections at two loops and beyond. A clear sign of such behaviour is the large uncertainty arising from residual dependence on the renormalization scale. Furthermore, in both of our benchmark points the ring-improved 1-loop effective potential fails to reproduce the nonperturbative critical temperature $T_c$ obtained from lattice simulations. We argue that this is due to higher-order resummations missing from the 1-loop potential.

On the contrary, resummations beyond the leading order are conveniently incorporated by performing dimensional reduction on the high-temperature theory. In section~\ref{sec:pert_theory} we show that the effective potential evaluated within the high-$T$ effective theory -- for which a 2-loop calculation is straightforward -- displays better convergence than its counterpart in the full theory and agrees with the lattice results well within accuracy. This result suggests that the aforementioned shortcomings of the resummed perturbation theory are not due to nonperturbative effects related to the ultrasoft thermal modes, but rather a consequence of bad convergence caused by large scalar couplings.

It should be pointed out that for cosmological applications, the thermodynamic quantities in Table~\ref{table:PT_comparison} should be obtained at the nucleation temperature $T_n$ instead of the critical temperature $T_c$. The computation of $T_n$ nevertheless requires precise knowledge of the critical temperature, which, according to our results, is beyond the reach of 1-loop resummed perturbation theory, and the 2-loop correction calculated in Ref.~\cite{Laine:2017hdk} does not significantly improve the result for $T_c$. For this reason, we advocate the use of the effective-theory approach even if the study is performed purely perturbatively, without numerical simulations. This strategy has already been recommended earlier in Refs.~\cite{Laine:2012jy,Laine:2017hdk}.

For both baryogenesis and gravitational-wave production, the phase transitions considered here are possibly too weak. A tempting resolution could be to move in the parameter space towards even larger couplings and consequently larger potential barriers -- a lapse that would come with a high price, as perturbative convergence will then only worsen. In fact, the accuracy of many results in recent literature which invoke large scalar couplings could be in jeopardy due to the 1-loop potential simply being too inaccurate in describing the phase transition. We believe that more care needs to be taken in such scenarios, if one is willing to push perturbation theory to its limits. Unfortunately, for very large couplings even the effective-theory approach is likely to fail, and purely nonperturbative methods would then be necessary. Naturally, one would expect this conclusion to also hold for other models which rely on large couplings for a strong phase transition.

Finally, we highlight an alternative setup for strong phase transitions where more elaborate dynamics of multiple light fields are responsible for strengthening the transition. While such multi-step transitions typically call for a degree of fine tuning, the requirement of large couplings could be avoided in these scenarios.

\section*{Acknowledgments}

This work was supported by the Academy of Finland grants 318319 and 308791. VK acknowledges the H2020-MSCA-RISE-2014 grant no. 645722 (NonMinimalHiggs). LN was supported by the Jenny and Antti Wihuri Foundation, TT by the Swiss National Science Foundation (SNF) under grant 200020-168988 and VV by the UK STFC Grant ST/P000258/1.  The lattice simulations were carried out on the Finnish IT Center for Science (CSC) supercomputers Taito and Sisu, and in part on the University of Helsinki cluster Kale (urn:nbn:fi:research-infras-2016072533). We are grateful to O. Gould, S. Huber, T. Konstandin, M. Laine, J. No, P. Schicho, A. Vuorinen and D.J. Weir for discussions, and thank G. Dorsch for providing the input parameters for the benchmark points in Table 4 of Ref.~\cite{Caprini:2015zlo}

\appendix

\section{Renormalization in the \MSbar scheme}
\label{sec:renormalization}

As mentioned in section~\ref{sec:inputs}, the input parameters for our analysis are the pole masses of the scalars $\{ M_h, M_H, M_A, M_{H^\pm} \}$, the scheme-dependent parameters $\{\tan\beta, \cos(\beta-\alpha),\mu^2\}$ and the gauge boson and fermion pole masses~\cite{Agashe:2014kda}. In order to relate these to parameters appearing in the Lagrangian, we apply a standard pole-mass renormalization at 1-loop level in the Minkowski space vacuum. As some of the couplings and masses in our theory are fairly large, loop effects can modify the relations between the two substantially. We emphasize that although this calculation is important for making accurate physical predictions, it is not directly related to the high-$T$ behavior of the theory, which is the main focus of our paper. 

Starting with the CP-conserving 2HDM, we first require that the VEVs, $v_i$ (c.f. Eq.~(\ref{eq:fields})), minimize the tree-level potential, 
\begin{equation}
\label{eq:min_cond}
    \frac{\partial V}{\partial \phi_i}\Big\vert_{\phi_i = \langle \phi_i \rangle}  = 0, \quad
    \frac{\partial V}{\partial \phi^\dagger_i}\Big\vert_{\phi_i = \langle \phi_i \rangle}  = 0,
\end{equation}
and rotate the fields to a diagonal basis as in Eq.~(\ref{eq:diagonalization}). The parameters $\{\lambda_{1-5},\mu^2_{11},\mu^2_{22}\}$ can be expressed in terms of the mass eigenvalues $m_i$, the mixing angles and $\mu^2$; the explicit relations can be found in {\em e.g.} Appendix B.2 of Ref.~\cite{Gorda:2018hvi}.

In the renormalized theory, dressed propagators are of the form 
\begin{align}
\label{eq:dressed_propagator}
G_i \sim \frac{1}{p^2 - m^2_i + \Pi_i(p^2,\Lambda)},
\end{align}
where $\Pi_i(p^2)$ denotes a self energy evaluated at external momentum $p$ and \MSbar scale $\Lambda$. Because the minimization conditions (\ref{eq:min_cond}) are imposed only at tree level, the VEVs generate one-particle-reducible tadpole contributions to the self energies that need to be accounted for, and are included in our study. The condition that the input masses $M_i$ correspond to poles of the loop-corrected propagators is then equal to having
\begin{align}
\label{eq:pole_cond}
m^2_i = M^2_i + \RE\Pi_i(M_i^2,\Lambda),
\end{align}
which is a system of four equations for the eigenvalues $\{m_h, m_H, m_A, m_{H^\pm}\}$. Similar pole-mass conditions are obtained for the gauge fields $W^\pm,Z$ as well as for the top quark $t$ (a detailed discussion can be found in Ref.~\cite{Kajantie:1995dw}). Pole conditions for the light fermions are not needed in our case, as their Yukawa couplings are negligible in the phase-transition analysis. 

Together with the direct input parameters $\tan\beta, \cos(\beta-\alpha)$ and $\mu^2$, these equations are sufficient to fix all but one parameter in the electroweak sector at some input scale $\Lambda_0$. The final relation is obtained from charged particle scattering in the Thomson limit. This fixes the electromagnetic fine-structure constant 
\begin{align}
\FShat = \frac{1}{4\pi}\frac{g^2 {g'}^2}{(g^2 + {g'}^2)}
\end{align}
in the \MSbar scheme via 
\begin{align}
\FShat\left( 1 + \frac{\delta\FS}{\FS} \right) \equiv \FShat\left( 1 + 2 \frac{g'}{g} \frac{\RE\Pi^{(T)}_{Z\gamma}(0)}{m_Z^2} + \RE\Pi'(0)_{\gamma\gamma} \right) = \FS,
\end{align}
where $\FS = 1/137.036$~\cite{Agashe:2014kda}, and the photon self energy $\Pi_{\gamma\gamma}$ and the (traverse) $Z\gamma$ correlation function are computed at one loop. In the 2HDM, $\delta\FS/\FS$ obtains a subdominant correction from $H^\pm$ loops. 

Using the above relations and omitting terms beyond leading order in the correlation functions, one obtains loop-corrected expressions for the renormalized parameters:
\begin{align}
\label{eq:vacuum-corrections}
\mu^2_{11}(\Lambda) ={}& \mu ^2 t_{\beta
   } + \frac{1}{2} \Big[ M_h^2 \Big( 1 + \frac{\RE\Pi _h\left(M_h^2,\Lambda \right)}{M_h^2}\Big) \left(c_{\beta -\alpha }^2+t_{\beta }
   c_{\beta -\alpha } s_{\beta -\alpha }-1\right) \nonumber \\
 &- M_H^2\Big( 1 + \frac{\RE\Pi _H\left(M_H^2,\Lambda \right)}{M_H^2}\Big) \left(c^2_{\beta -\alpha }+t_{\beta } c_{\beta -\alpha }s_{\beta -\alpha }\right)\Big] , \\
\mu^2_{22}(\Lambda) ={}& \mu^2 t^{-1}_\beta + \frac12 \Big[ M_h^2 \Big( 1 + \frac{\RE\Pi _h\left(M_h^2,\Lambda \right)}{M_h^2}\Big)(c_{\beta -\alpha }^2 - t^{-1}_{\beta } c_{\beta -\alpha } s_{\beta -\alpha }-1) \nonumber \\
& - M_H^2\Big( 1 + \frac{\RE\Pi _H\left(M_H^2,\Lambda \right)}{M_H^2}\Big) (c^2_{\beta -\alpha }-t^{-1}_{\beta } c_{\beta -\alpha }s_{\beta -\alpha }) \Big] , \\
\lambda_1(\Lambda) ={}& \frac{\pi\FS M_Z^2}{M_W^2(M_Z^2-M_W^2)} \frac14 c^{-2}_{\beta} \Big[ \Big( -2\mu^2 t_\beta + M_h^2 + M_H^2 - (M_h^2 - M_H^2)c_{2\alpha} \Big) \nonumber \\
&\times \Big( 1 - \frac{\delta\FS}{\FS} + \frac{\RE\Pi^{(T)}_W(M_W^2, \Lambda)}{M_W^2} - \frac{\RE\Pi^{(T)}_Z(M_Z^2, \Lambda)}{M_Z^2} \nonumber \\ 
&+ \frac{\RE\Pi^{(T)}_Z(M_Z^2, \Lambda) - \RE\Pi^{(T)}_W(M_W^2, \Lambda)}{M_Z^2 - M_W^2} \Big) + 2s^2_\alpha \RE\Pi_h(M_h^2,\Lambda) \nonumber \\
&+ 2 c^2_\alpha \RE\Pi_H(M_H^2,\Lambda) \Big] ,\\
\lambda_2(\Lambda) ={}& \frac{\pi\FS M_Z^2}{M_W^2(M_Z^2-M_W^2)} \frac14 s^{-2}_{\beta} \Big[  \Big( -2\mu^2 t^{-1}_\beta + M_h^2 + M_H^2 + (M_h^2 - M_H^2)c_{2\alpha} \Big) \nonumber \\
&\times \Big( 1 - \frac{\delta\FS}{\FS} + \frac{\RE\Pi^{(T)}_W(M_W^2, \Lambda)}{M_W^2} - \frac{\RE\Pi^{(T)}_Z(M_Z^2, \Lambda)}{M_Z^2} \nonumber \\ 
&+ \frac{\RE\Pi^{(T)}_Z(M_Z^2, \Lambda) - \RE\Pi^{(T)}_W(M_W^2, \Lambda)}{M_Z^2 - M_W^2} \Big) + 2c^2_\alpha \RE\Pi_h(M_h^2,\Lambda) \nonumber \\
&+ 2 s^2_\alpha \RE\Pi_H(M_H^2,\Lambda) \Big] , \\
\lambda_3(\Lambda) ={}& \frac{\pi\FS M_Z^2}{M_W^2(M_Z^2-M_W^2)} \Big[ s^{-1}_{2\beta} \Big( -2\mu^2 -(M_h^2 - M_H^2)s_{2\alpha} + 2 M_{H^\pm}^2 s_{2\beta} \Big) \Big( 1 - \frac{\delta\FS}{\FS} \nonumber \\
&+ \frac{\RE\Pi^{(T)}_W(M_W^2, \Lambda)}{M_W^2} - \frac{\RE\Pi^{(T)}_Z(M_Z^2, \Lambda)}{M_Z^2} + \frac{\RE\Pi^{(T)}_Z(M_Z^2, \Lambda) - \RE\Pi^{(T)}_W(M_W^2, \Lambda)}{M_Z^2 - M_W^2} \Big) \nonumber \\
&+ \big(\RE\Pi_h(M_h^2,\Lambda) - \RE\Pi_H(M_H^2,\Lambda)\big) ( s_{\beta-\alpha} + c_{\beta-\alpha}t^{-1}_\beta )( s_{\beta-\alpha} - c_{\beta-\alpha}t_\beta ) \nonumber \\ 
&+ 2\RE\Pi_{H^\pm}(M_{H^\pm}^2,\Lambda)\Big] ,\\
\lambda_4(\Lambda) ={}& \frac{\pi\FS M_Z^2}{M_W^2(M_Z^2-M_W^2)} \Big[ \Big( \mu^2 s^{-1}_\beta c^{-1}_\beta + M_A^2 - 2M_{H^\pm}^2  \Big) \Big( 1 - \frac{\delta\FS}{\FS} + \frac{\RE\Pi^{(T)}_W(M_W^2, \Lambda)}{M_W^2} \nonumber \\
& - \frac{\RE\Pi^{(T)}_Z(M_Z^2, \Lambda)}{M_Z^2} + \frac{\RE\Pi^{(T)}_Z(M_Z^2, \Lambda) - \RE\Pi^{(T)}_W(M_W^2, \Lambda)}{M_Z^2 - M_W^2} \Big) \nonumber \\
&+ \RE\Pi_A(M_A^2,\Lambda) - 2\RE\Pi_{H^\pm}(M_{H^\pm}^2,\Lambda)\Big], \\
\lambda_5(\Lambda) ={}& \frac{\pi\FS M_Z^2}{M_W^2(M_Z^2-M_W^2)} \Big[ \Big( \big( t_\beta + t^{-1}_\beta \big)\mu^2 - M_A^2 \Big) \Big( 1 - \frac{\delta\FS}{\FS} + \frac{\RE\Pi^{(T)}_W(M_W^2, \Lambda)}{M_W^2} \nonumber \\
& - \frac{\RE\Pi^{(T)}_Z(M_Z^2, \Lambda)}{M_Z^2} + \frac{\RE\Pi^{(T)}_Z(M_Z^2, \Lambda) - \RE\Pi^{(T)}_W(M_W^2, \Lambda)}{M_Z^2 - M_W^2} \Big) - \RE\Pi_A(M_A^2, \Lambda)\Big] , \\
g^2(\Lambda) ={}&  \frac{4\pi\FS M_Z^2}{M_Z^2 - M_W^2}\Big[1-\frac{\delta\FS}{\FS} - \frac{\RE\Pi^{(T)}_Z(M_Z^2,\Lambda)}{M_Z^2} + \frac{\RE\Pi^{(T)}_Z(M_Z^2,\Lambda)-\RE\Pi^{(T)}_W(M_W^2,\Lambda)}{M_Z^2-M_W^2}\Big] , \\
{g'}^2(\Lambda) ={}&  \frac{4\pi\FS M_Z^2}{M_W^2}\Big[1-\frac{\delta\FS}{\FS} - \frac{\RE\Pi^{(T)}_Z(M_Z^2,\Lambda)}{M_Z^2} + \frac{\RE\Pi^{(T)}_W(M_W^2,\Lambda)}{M_W^2}\Big] , \\
\label{eq:top-yukawa}
y_t^2(\Lambda) ={}&  \frac{2\pi \FS M_Z^2 M_t^2}{s^2_\beta M_W^2(M_Z^2 - M_W^2)}\Big[1 - \frac{\delta\FS}{\FS} - \frac{\Pi^{(T)}_Z(M_Z^2,\Lambda)}{M_Z^2} + \frac{\Pi^{(T)}_W(M_W^2,\Lambda)}{M_W^2} \nonumber \\
 & + \frac{\Pi^{(T)}_Z(M_Z^2,\Lambda)-\Pi^{(T)}_W(M_W^2,\Lambda)}{M_Z^2-M_W^2} - 2\big(\Sigma_s(M_t^2,\Lambda) + \Sigma_v(M_t^2,\Lambda)\big)\Big],
\end{align}
where $t_\beta \equiv \tan\beta$. In Eq.~(\ref{eq:top-yukawa}), $\Sigma_s$ and $\Sigma_v$ are the scalar and vector parts of the top quark self energy, and the axial and axial vector parts do not enter the pole-mass condition. As in Refs.~\cite{Kajantie:1995dw,Laine:2017hdk,Niemi:2018asa}, we neglect loop corrections to the $\gr{SU(3)}$ coupling $g_s$ and fix its value at tree level.

The expressions for the self energies are fairly long and, for the sake of readability, will not be listed here. Explicit formulas can however be found in Ref.~\cite{Kanemura:2015mxa}, and we have verified that our results match the expressions in their Appendix C. In practice, we have used the Feynman-t'Hooft gauge to simplify the calculation of the self energies. To 1-loop order, the UV counterterms required for renormalization are the same as those used in dimensional reduction. Apart from gauge-dependent terms, the counterterms can be found in Ref.~\cite{Gorda:2018hvi}. 

Eqs.~(\ref{eq:vacuum-corrections})-(\ref{eq:top-yukawa}) form a non-linear system of equations for the renormalized parameters, as the correlations functions and $\delta\FS/\FS$ themselves are functions of the \MSbar couplings. The situation can, however, be simplified by substituting the renormalized parameters with the physical, scheme-independent parameters inside the loop corrections, {\em i.e}, making the replacements $m_i(\Lambda) \rightarrow M_i$ and $\FShat(\Lambda) \rightarrow \FS$, and the difference between the two prescriptions is formally of higher order in perturbation theory. 

In the IDM limit ($\mu^2 = 0$ and $v_1=0$), the calculation proceeds analogously, but instead of $\tan\beta$ and $\cos(\beta-\alpha)$ we input the self-coupling $\lambda_1(\Lambda)$ and the combination $\big(\lambda_3(\Lambda)+\lambda_4(\Lambda)+\lambda_5(\Lambda)\big)/2$. Detailed formulas in the IDM case (neglecting $g'$ contributions to the self energies) can be found in Ref.~\cite{Laine:2017hdk}. In Ref.~\cite{Laine:2017hdk}, it is also discussed how higher-order corrections to Eqs.~(\ref{eq:vacuum-corrections})-(\ref{eq:top-yukawa}) can be partially resummed by solving the equations ``self-consistently'', without performing the linearization. In BM1, we adopt their approach and solve the parameters iteratively, dropping $g'$ terms inside the loop corrections. This ensures that our study of thermal effects in BM1 is directly comparable to the 2-loop results of Ref.~\cite{Laine:2017hdk}.

In BM2 however, the iterative approach does not converge very well, whereby we take the simplified approach. We find corrections of order $10\%$ to the couplings relative to their tree-level values. For the smallest coupling $\lambda_2$, however, the correction is $\sim 25\%$, while the mass parameter $\mu^2_{22}$ is modified by $\sim 50\%$. Such large corrections, and the bad convergence of the iterative approach, again indicate that our BM2 is already close to the border of applicability of perturbation theory.

\bibliographystyle{JHEP}

\bibliography{2HDM_lattice}

\end{document}